\documentclass[preprint,aps,superscriptaddress,showpacs,citeautoscript]{revtex4-1}
\UseRawInputEncoding
\usepackage{graphicx}
\usepackage{dcolumn}
\usepackage{bm}
\usepackage{flafter}
\usepackage{float}
\usepackage{lettrine}
\usepackage{amsmath}
\usepackage{amssymb}
\usepackage{color}
\usepackage{epsfig}
\graphicspath{{eps+pdf/}}

\begin{document}

\date{\today}

\title{Structural study of TATB  under detonation-induced shock conditions}

\author{Elissaios Stavrou}
\email{elissaios.stavrou@gtiit.edu.cn}
\affiliation {Lawrence Livermore National Laboratory, Physical and Life Sciences Directorate,
Livermore, California 94550, USA}
\affiliation{Materials and Engineering Science Program, Guangdong Technion-Israel Institute of Technology, Shantou, Guangdong, 515063, China}
\affiliation {Technion-Israel Institute of Technology, Haifa, 32000, Israel}
\author{Michael Bagge-Hansen}
\affiliation {Lawrence Livermore National Laboratory, Physical and Life Sciences Directorate,
Livermore, California 94550, USA}
\author{Joshua A. Hammons}
\affiliation {Lawrence Livermore National Laboratory, Physical and Life Sciences Directorate,
Livermore, California 94550, USA}
\author{Michael H.  Nielsen}
\affiliation {Lawrence Livermore National Laboratory, Physical and Life Sciences Directorate,
Livermore, California 94550, USA}
\author{William L. Shaw}
\affiliation {Lawrence Livermore National Laboratory, Physical and Life Sciences Directorate,
Livermore, California 94550, USA}
\author{Will Bassett}
\affiliation {Lawrence Livermore National Laboratory, Physical and Life Sciences Directorate,
Livermore, California 94550, USA}
\author{Thomas W. Myers}
\affiliation {Lawrence Livermore National Laboratory, Physical and Life Sciences Directorate,
Livermore, California 94550, USA}
\author{Lisa M. Lauderbach}
\affiliation {Lawrence Livermore National Laboratory, Physical and Life Sciences Directorate,
Livermore, California 94550, USA}
\author{Ralph L. Hodgin}
\affiliation {Lawrence Livermore National Laboratory, Physical and Life Sciences Directorate,
Livermore, California 94550, USA}
\author{Nicholas A. Perez-Marty}
\affiliation {Lawrence Livermore National Laboratory, Physical and Life Sciences Directorate,
Livermore, California 94550, USA}
\author{Kamel Fezzaa}
\affiliation {Advanced Photon Source, Argonne National Laboratory, 9700 South Cass Avenue, Argonne, IL 60439, USA}
\author{Alex Deriy}
\affiliation {Advanced Photon Source, Argonne National Laboratory, 9700 South Cass Avenue, Argonne, IL 60439, USA}
\author{Sorin Bastea}
\affiliation {Lawrence Livermore National Laboratory, Physical and Life Sciences Directorate,
Livermore, California 94550, USA}
\author{Laurence E. Fried}
\affiliation {Lawrence Livermore National Laboratory, Physical and Life Sciences Directorate, Livermore, California 94550, USA}
\author{Lara D. Leininger}
\affiliation {Lawrence Livermore National Laboratory, Physical and Life Sciences Directorate,
Livermore, California 94550, USA}
\author{Trevor M. Willey}
\email{willey1@llnl.gov}
\affiliation {Lawrence Livermore National Laboratory, Physical and Life Sciences Directorate,
Livermore, California 94550, USA}

\begin{abstract}
We explore the response of the  insensitive high explosive (IHE) 1,3,5-Triamino-2,4,6-trinitrobenzene (TATB)   under detonation-induced shock conditions using $in-situ$ synchrotron X-ray diffraction in the100 ns time-scales, using either a conventional or a colliding detonation drive. In all of the detonation experiments on various sizes and morphologies  of TATB, we observe an extended stability of the TATB  crystal structure.  As the detonation front passes through TATB, X-ray diffraction indicates a portion of the TATB exhibits a compression up to 30+ GPa, followed subsequently by a pressure release and continued decomposition over a few hundred nanoseconds. Likewise, for colliding detonation-driven shock compression of single crystals of TATB, a significant portion of crystalline TATB  appears to be stable up to 60+ GPa. Conversely, in similar detonations of an LLM-105 polymer-bonded explosive, X-ray diffraction is simply indicative of fast decomposition without the apparent compression and slow decomposition seen in TATB.  The results indicate the surprising resilience of TATB under these high-pressure, temperature and shock conditions, providing a baseline for understanding the insensitivity of TATB.  The results also provide intriguing information for the extended reaction zone in TATB, and the hot-spot mechanisms for initiating and propagating detonation in this uniquely insensitive explosive.
\end{abstract}

\maketitle

\section{Introduction}
Energy release in detonation and insensitivity to hazards are the two most important parameters  that determine the effectiveness and safety of a high explosive (HE). Given  that these two characteristics are usually incompatible,   1,3,5-Triamino-2,4,6-trinitrobenzene (TATB),  stands out as the optimum choice when insensitivity/safety is of utmost importance. Indeed, among similar materials with comparable explosive energy release, TATB is remarkably difficult to shock-initiate, has a low friction sensitivity, and is thermally stable at ambient pressure to approximately 615 K \cite{Rice1990}. Thus, probing the structural/molecular evolution of TATB under detonation conditions is of key importance towards understanding the origin of its insensitivity.  For these reasons, TATB has attracted extensive research effort aiming to elucidate its high-pressure structural behavior (up to and beyond detonation pressures, nominally 30-40 GPa \cite{Tarver2006}) under both static and dynamic compression.

Under static compression and according to a plethora of previous theoretical studies \cite{Manaa2012,Rykounov2015,Bedrov2009,Byrd2007,Wu2014,Landerville2010,Qin2019,Qin2019,Budzevich2010,Fedorov2014,Kroonblawd2020},  TATB remains in the ambient pressure crystal structure up to at least 100 GPa. Moreover, experimental isothermal high-pressure powder and polycrystalline X-ray diffraction (XRD) studies \cite{Stevens2008,Plisson2017} do not report any structural phase transition up to 70 GPa. Very recently, a subtle second order phase transition towards a monoclinic crystal structure  was reported above 4 GPa \cite{Steele2019a}. Thus, the consensus in the field is that under static compression TATB retains its basic molecular/structural motif even above detonation pressures.

In contrast to the case of static compression, the structural evolution of TATB under detonation/shock conditions and the possibility of chemical reactions under these high-pressure and high-temperature conditions remains an open question. This is, in part,  due to the lack of $in-situ$ X-ray diffraction (XRD) measurements under such conditions. Previous shock Hugoniot measurements of pure TATB (powder or single crystal)  and TATB-based polymer-bonded explosives  (PBX) have reported the shock Hugoniot equation of state (EOS) up to 83 GPa \cite{Gustavsen2006,Dick1988,Jackson1976,Dallman1993,Marshall2020}. However, in these studies only the density as a function of pressure was determined under shock conditions.

To address this issue, we have performed an $in-situ$ synchrotron XRD  study of ultrafine powders of  TATB (UFTATB), single crystals of TATB (SCTATB) and of a TATB-based PBX, under detonation-induced shock  conditions at 100 ns time scales. Aiming  to probe larger quantities, and thus increase the confidence of the Bragg peaks assignments  of  TATB under shock conditions,  we use detonation to shock macroscopic quantities of TATB  in a  geometry that X-rays are orthogonal to the shock front. This allowed us to probe the crystal structure and confirm the structural stability of the ambient-conditions triclinic phase  of TATB under detonation-induced shock conditions up to 60+ GPa and also during pressure release towards ambient pressure. In contrast to the TATB-based specimens, identical experimental runs using an LLM-105 (2,6-diamino-3,5-dinitropyrazine-1-oxide, C$_4$H$_4$N$_6$O$_5$, Lawrence Livermore Molecule No.105) based PBX  did not exhibit diffraction that might indicate detonation induced compression of the respective  ambient-conditions crystal structures.

\section{Methods}

High purity UFTATB (with a $<$ 5 $\mu$m nominal median particle size  \cite{Nandi2007,Foltz1996}), SCTATB \cite{Han2009} and  a TATB-based PBX (nominally 95\% TATB, with a broad 5-90 $\mu$m particle size distribution \cite{Peterson2005})  of different sizes were  used for all XRD experiments. Moreover, LLM-105 based PBX (nominally 96\% LLM-105) specimens were also used aiming to compare with the structural stability of TATB. Diffraction experiments are limited to gram quantities due to the experimental setup, geometry, and facility.  Due to these limitations, samples were boosted.   UFTATB as well as TATB and LLM-105 PBXs were generally 1/4" (6.35 mm) in diameter, and about 2.0 mm tall. Two different PBXs were used as boosters, one was based on hexanitrohexaazaisowurtzitane  (CL-20, the other based on pentaerythritol tetranitrate (PETN), to generate different peak drive pressures. The experiments with both boosters produced similar results with only different peak pressures. For simplicity we only present the results obtained by using CL-20 based boosters. These CL-20 or PETN based boosters are each about about 0.22" (5.6 mm) tall.  UFTATB and TATB and LLM-105 PBXs  samples were either placed on top of a single PBX or sandwiched between two PBXs that have the same radius with the TATB specimens, see fig. 1. The SCTATB specimens were planar, with a sub-mm thickens and the c-axis roughly parrel to the thickness (thinner) dimension, and irregularly shaped in the lateral dimensions. The maximum lateral dimensions were smaller than the  diameters of the boosters. This way the peak shock pressure was controlled by both the type of HE (30-40 GPa peak pressure) and also by what we will refer to as  \textquotedblleft conventional\textquotedblright {}  or \textquotedblleft colliding\textquotedblright {} detonation.  A few 3/8" (9.53 mm) diameter, 11.57 mm tall, 1.5 g UFTATB shots were also fired and boosted by a 3.9 mm tall  PBX.

\begin{figure}[ht]
{\includegraphics[width=\linewidth]{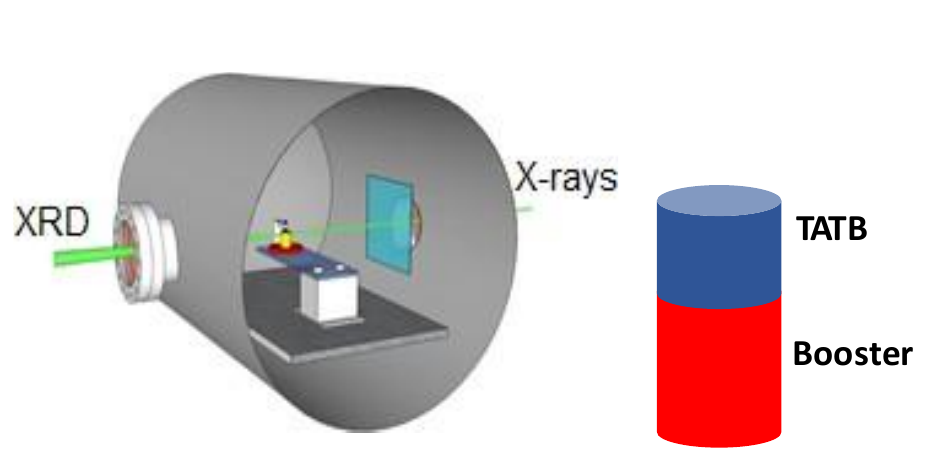}}
\centering
\caption{Schematic diagram of the detonation tank  for in-situ XRD under detonation conditions and schematic section (not to scale) of the experimental cylindrical setup showing the TATB  cylinder on top of a HE cylinder. The cylindrical setup is aligned in a way that the incident X-ray beam is at the center (vertical and lateral) of the TATB cylinder and  perpendicular to the shock front. }
\end{figure}

Time-resolved XRD measurements were performed within a Lawrence Livermore National Laboratory (LLNL) detonation tank at the Advanced Photon Source (APS), Argonne National Laboratory, at the Dynamic Compression Sector, within the special purpose hutch (35ID-B)\cite{Gustavsen2006,Bagge-Hansen2015,Watkins2017} or at 32ID-B \cite{Willey2016}.  The explosive sample assemblies were placed within a 120L steel vacuum vessel (Teledyne RISI) and pumped down to $<$ 200mTorr, see Fig. 1. The tank uses upstream and downstream Kapton$^{TM}$(polyimide) windows to facilitate the X-ray transmission geometry required for XRD and/or radiography under low vacuum conditions. Within the vacuum vessel, Lexan$^{TM}$ (polycarbonate) panels were used as shrapnel mitigation. The samples were detonated near the rear window to increase the angular range of the detector. A Tantalum beamstop was placed between two 2 mm polycarbonate plates that were placed a few cm from the specimens. More details about the LLNL detonation tank and the experimental setup can be found in Refs. \onlinecite{Bagge-Hansen2019,Hammons2019}.  The specimens were aligned in the beam such that the focused beam was centered horizontally and vertically on the 2.0 mm 1/4" diameter samples (1.0 mm above the booster-sample interface.) The beam was simply centered vertically, and roughly horizontally on single-crystal samples.  The x-ray beam was aligned 2.0 mm below the top surface of the larger 3/8" samples. The X-ray beam had an asymmetric energy spectrum, with a peak near 24.3 keV and a bandwidth of $\sim$ 1 keV \cite{Wang2019}.

Detonation is synchronized with the APS bunch clock, thus permitting XRD from discrete 34ps rms X-ray pulses, which arrive every 153.4 ns during 24-bunch mode. The sample to detector distance was about 11cm. Scattering intensity was recorded using an array of four identical area detectors (PI-MAX4 1024i ICCD, Princeton Instruments) focused on the output of an image intensifier \cite{Bagge-Hansen2015,Jensen2014,Gupta2012,Turneaure2016}. The image intensifier (Photek, MCP140) is equipped with a P47 phosphor and a LSO:Ce scintillator (DMI/Reading Imaging) that have decay times of 80 ns and 40 ns, respectively. Each of these cameras have \textquotedblleft Dual Image Feature\textquotedblright (DIF) capability that allows acquisition of two image frames spaced at least 500 ns apart. This way, a total of eight frames (noted as frames C1-C8 and separated by 153.4 ns) are recorded for each detonation event. Given that XRD images are rerecorded every 153.4 ns, a $\thickapprox$10\% of the signal recorded in a given frame (XRD image) is expected to originate from the residual XRD image of the previously recorded frame e.g. in frame C3 a $\thickapprox$10\% residual intensity of frame C2 is recorded. The XRD intensity from different cameras (C1-C4) are normalized in intensity using the static images acquired on each camera immediately before each detonation event. Aiming to acquire  XRD images in time intervals shorter than  153.4 ns, we used different shots of identical setups, \emph{i.e.} both HEs under study and boosters were identical in the relevant shots. Aiming to ensure repeatability, we tested the identical setups under, at least, 2-3 shots with the same timing. In all shots the results were practically identical, with any difference in respect of time (shock front arrival) and  Bragg peaks positions smaller than our experimental resolution.

Integration of powder diffraction patterns to yield scattering intensity versus 2$\theta$ diagrams and initial analysis were performed using the DIOPTAS program \cite{Prescher2015}. DIOPTAS was also used to remove/subtract the intensity contribution of the previous frame discussed above.   For each frame (XRD image) the previous frame was used as the background image scaled to the point that the corresponding residual diffraction intensity disappears. A 15-20\% scaling was sufficient to, almost, totally remove the residual intensity. This slightly higher scaling compared to what is expected due to the experimental setup (residual intensity), is attributed to parts of the specimens at the edge/surface of the cylinders that are at ambient conditions even under detonating-induced shock conditions \cite{Stavrou2020}.  Si640E (NIST standard) and CeO$_2$ were used as calibrants for the XRD sample-detector geometry. Calculated XRD patterns were produced using the POWDER CELL program \cite{Kraus1996}, for the corresponding crystal structures  assuming continuous Debye rings of uniform intensity. Indexing of XRD patterns was performed using the DICVOL program \cite{Boutlif2004} as implemented in the FullProf Suite. The X-ray wavelength was $\lambda$=0.52\AA.

\section{Results}

Figure 2 shows integrated diffraction patterns of the TATB PBX and SCTATB  at selected times after the single detonation front reaches the probing spot (t=0). XRD patterns with residual intensity subtraction are shown in Supporting material Fig. S1 \cite{supp}.  In the case of the PBX, the observed  diffraction pattern (noted as pre-shock) of TATB  before the detonation front reaches the probing spot  in in good agreement with the calculated pattern of TATB \cite{Cady1965}, after taking into account the spectral characteristics of the incident X-ray beam. This implies that the grain size of TATB in the PBX is much smaller than the X-ray beamsize (nominally 20 $\mu$m X50 $\mu$m) and resembles  a powder-like TATB specimen; this is further supported by the 2D X-ray diffraction image (see Fig. 2(c)) that exhibits apparent Debye rings of uniform intensity. On the other hand, a normally expected difference can be seen between the observed and calculated pattern in the case of SCTATB.

\begin{figure}[ht]
{\includegraphics[width=\linewidth]{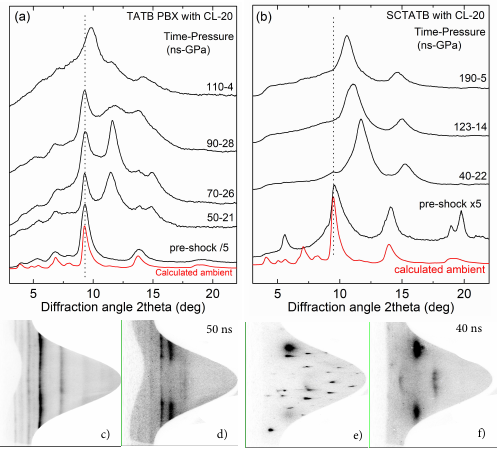}}
\centering
\caption{ X-ray diffraction patterns,   of a)the TATB PBX and b) SCTATB under detonation induced, using a CL-20 based PBX, shock conditions at selected times. The calculated, after taking into account the spectral characteristics of the incident x-ray beam  pattern of triclinic TATB \cite{Cady1965} at ambient conditions   is plotted with red. The vertical dashed line indicates the position of the most intense 002 Bragg peak  of TATB at ambient conditions. The static patterns in a) and b) are scaled by /5 and X5, respectively for clarity. 2D X-ray diffraction images in rectangular coordinates (cake) below the patterns correspond to the static (left) and under shock conditions (right) diffraction patterns of the TATB PBX (c) and d)) and SC-TATB (e) and f)), respectively. The maximum pressure is for TATB PBX and SCTATB is, 28 and 22 GPa, respectively.  }
\end{figure}

Both the TATB PBX and SCTATB diffraction patterns exhibit a clear pressure induced shift of the most intense TATB peaks, located at  9.29$^o$ (002) and at 14.08$^o$  (convolution of TATB Bragg reflections, see Fig. S2)  as a function of time. We note the much lower intensity of the pre-shock SCTATB pattern, because the X-ray beam   was roughly orthogonal to the c-axis,  in comparison to the one of TATB PBX. For this reason, the residual XRD intensity from the ambient pattern of SCTATB can be barely observed at later times while, it is clearly observed in the case of TATB PBX.   It is noteworthy that, as evident by the cake images ( Fig. 2(e) and (f)), the morphology of the SCTATB is substantially altered under shock conditions, evolving towards a multigrain morphology with a more uniform and enhanced diffraction intensity distribution. Interestingly, the main TATB Bragg peaks of both PBX and the SCTATB exhibit a pressure-induced shift that evolves to higher angles but then evolves back to lower angles as the   shock pressure releases at later 100-ns timescales as the detonation products recover  towards ambient pressure. Thus, part of the TATB sample retains  crystallinitye  during the whole detonation-induced shock event. The absence of any observable phase change of TATB might suggest that the triclinic structure is indeed stable \cite{Stevens2008,Plisson2017}  up to molecular breakdown; however, the recently reported monoclinic HP \cite{Steele2019a} structure is indistinguishable from the triclinic structure in our data. So, a portion of the TATB  not only withstand the peak pressures and temperatures  created by the detonation but also remains during the pressure release towards the ambient conditions. This observation is probably related to the fact  that  TATB is remarkably difficult to shock-initiate.

As it is apparent in Fig. 2(a), the intensity of the TATB Bragg peaks under shock conditions are much less intense than the peaks  in the pre-shock pattern. A more detailed analysis, based on the relative integrated intensities of the 002 Bragg peak and the convolution of peaks at $\thickapprox$ 14$^o$ (at ambient conditions), revealed that up to $\thickapprox$30\%  of  initial crystalline TATB remains at 50 ns. It is plausible to assume that the at least 70\% of the initial  TATB has undergone a chemical transformation towards a non-crystalline state. This is further supported by the clear increase of the broad background   (between roughly  7-17 $^o$) that resembles the XRD expected for non-crystalline materials (amorphous or liquids/gases). Thus, the recorded XRD pattern represents the superposition of TATB that has undergone a chemical reaction plus approximately 30\% of crystalline TATB under pressure in areas of the specimens where TATB crystallites persist.

To test even higher-pressure detonation conditions, figure 3 shows integrated diffraction patterns of SCTATB  at selected times after the colliding detonation-driven shock front reaches the probing spot (t=0). XRD patterns with residual intensity subtraction are shown in  Fig. S3. Similarly with the conventional detonation, the morphology of SCTATB evolves towards a multigrain morphology under shock conditions. The most   intense TATB peaks,  at ambient conditions, can be observed up to the maximum pressure achieved using the colliding setup. Thus,  a portion of the TATB remains a crystalline material,  even far above the conventional detonation pressures (30-40 GPa) and the recently predicted \cite{Hamilton2019} 14-34 GPa threshold for detonation.

\begin{figure}[ht]
{\includegraphics[width=\linewidth]{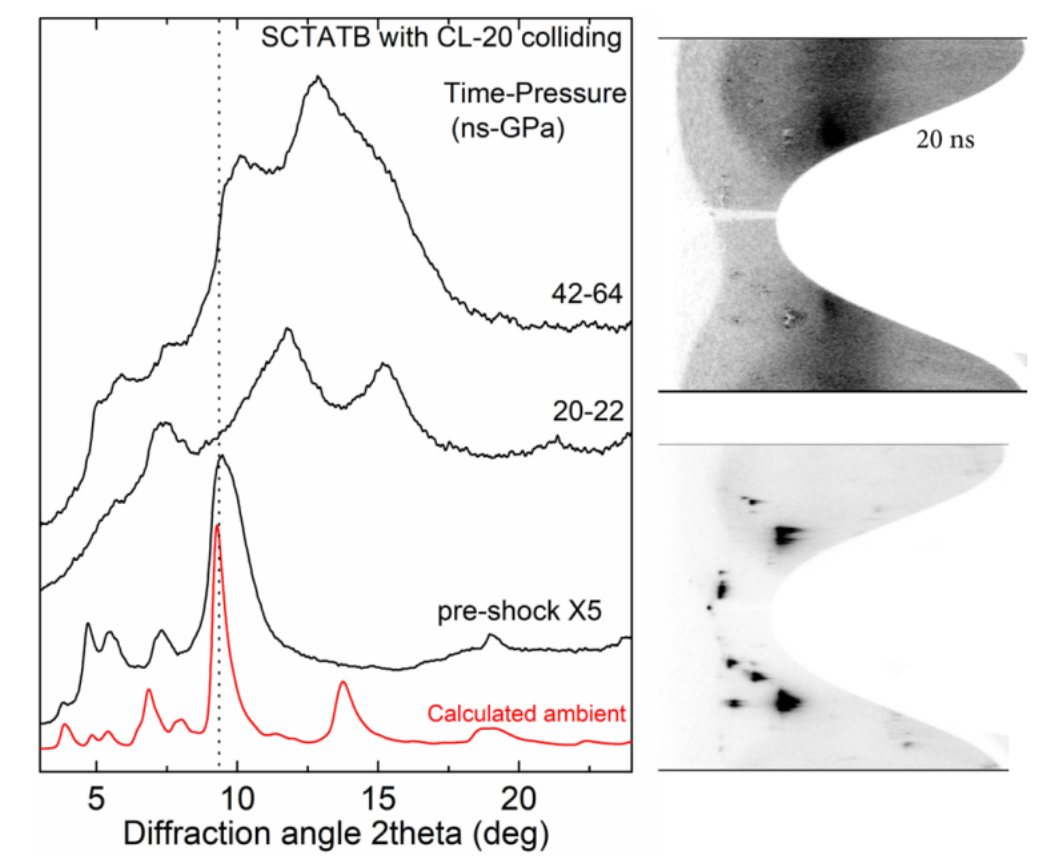}}
\centering
\caption{ X-ray diffraction patterns of SCTATB  under detonation-induced shock compression, using a CL-20 based PBX in colliding arrangement,  at selected times. The calculated pattern of triclinic TATB \cite{Cady1965} at ambient conditions   is plotted with red. The vertical dashed line indicates the position of the most intense 002 Bragg peak  of TATB at ambient conditions. The static patterns is scaled by X5  for clarity. Times are relative to the detonation front passing through the X-ray beam.  2D X-ray diffraction images in rectangular coordinates (cake) at the right of  the patterns correspond to the static  and under shocked condition diffraction patterns  of SCTATB, respectively. The maximum pressure is 64 GPa.}
\end{figure}

We estimated the corresponding pressures as a function of time by using the known static room temperature (RT)  EOSs of TATB \cite{Plisson2017,Steele2019a} \emph{i.e.} by comparing the $d$-spacing position of the observed 002 Bragg peak (by far the most intense Bragg peak of TATB)  with the expected position at a given pressure  according to the previously reported isothermal RT EOSs. More details about the pressure estimation are given in  Fig. S2. Given that we expect elevated temperatures during the detonation-induced shock compression, the estimated pressures correspond to the lower limit of the actual pressures.  We note that in previous studies (e.g.  Refs. \onlinecite{Leversee2019,Degtyarev2016}) a good agreement between the  EOSs of TATB under static and shock compression was observed up to, at least, 20 GPa.  The results  are summarized in Fig.4. In the case of conventional detonation-driven shock compression, for all TATB samples we observe practically the same peak pressure (25-27 GPa) in agreement with what is expected for a lower bound in TATB detonation pressures where the von Neumann spike is estimated to be about 34 GPa and Champman-Jouget is 25-27 GPa \cite{Tarver2006}. In our measurements t=0 is defined as the time the shock front reaches the X-ray position probing spot, that is nominally the center of the specimen. Thus, the time we observe the peak pressure is mainly defined by the corresponding height of the specimen. Nevertheless, all TATB specimens exhibit a similar durations of the shock effect of $\approx$75-100 ns. Likewise, for the colliding-driven shock compression of SCTATB, a peak pressure of $\approx$ 60 GPa is determined with a $\approx$75-100 ns shock duration.

\begin{figure}[ht]
{\includegraphics[width=\linewidth]{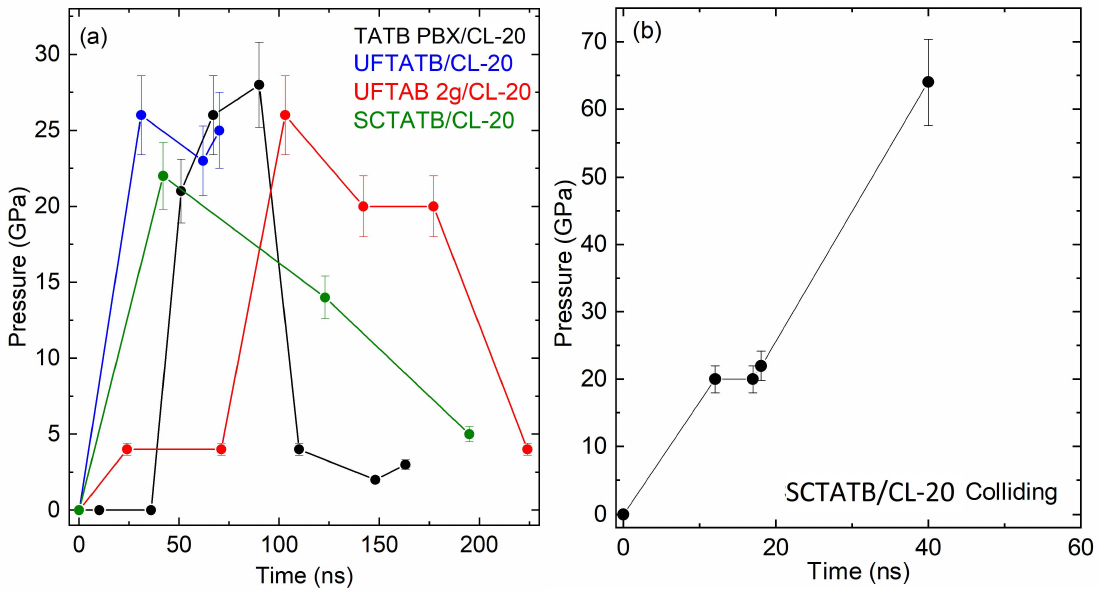}}
\centering
\caption{Pressure as a function of time  for a) single and b) colliding detonation driven shock compression of the various TATB specimens used in this study.}
\end{figure}

To compare with TATB-based specimens another relatively insensitive explosive, an  LLM-105-based PBX  was also examined under the same conventional detonation conditions. Integrated diffraction patterns of the LLM-105 PBX, at selected times after the conventional detonation front reaches the probing spot (t=0), are shown in Figure 5. Reasonable,  agreement is observed between the diffraction pattern (noted as pre-shock) of this PBX  before the detonation front reaches the probing spot    and the calculated pattern of LLM-105 \cite{Gilardi2001}.  In contrast to the case of the TATB PBX   the 2D X-ray diffraction image, see inset of Fig. 5, reveals the presence of spotty-like  Debye rings of non-uniform intensity.This implies that the grain size of LLM-105 in the PBX  is comparable to the  X-ray beamsize.

\begin{figure}[ht]
{\includegraphics[width=\linewidth]{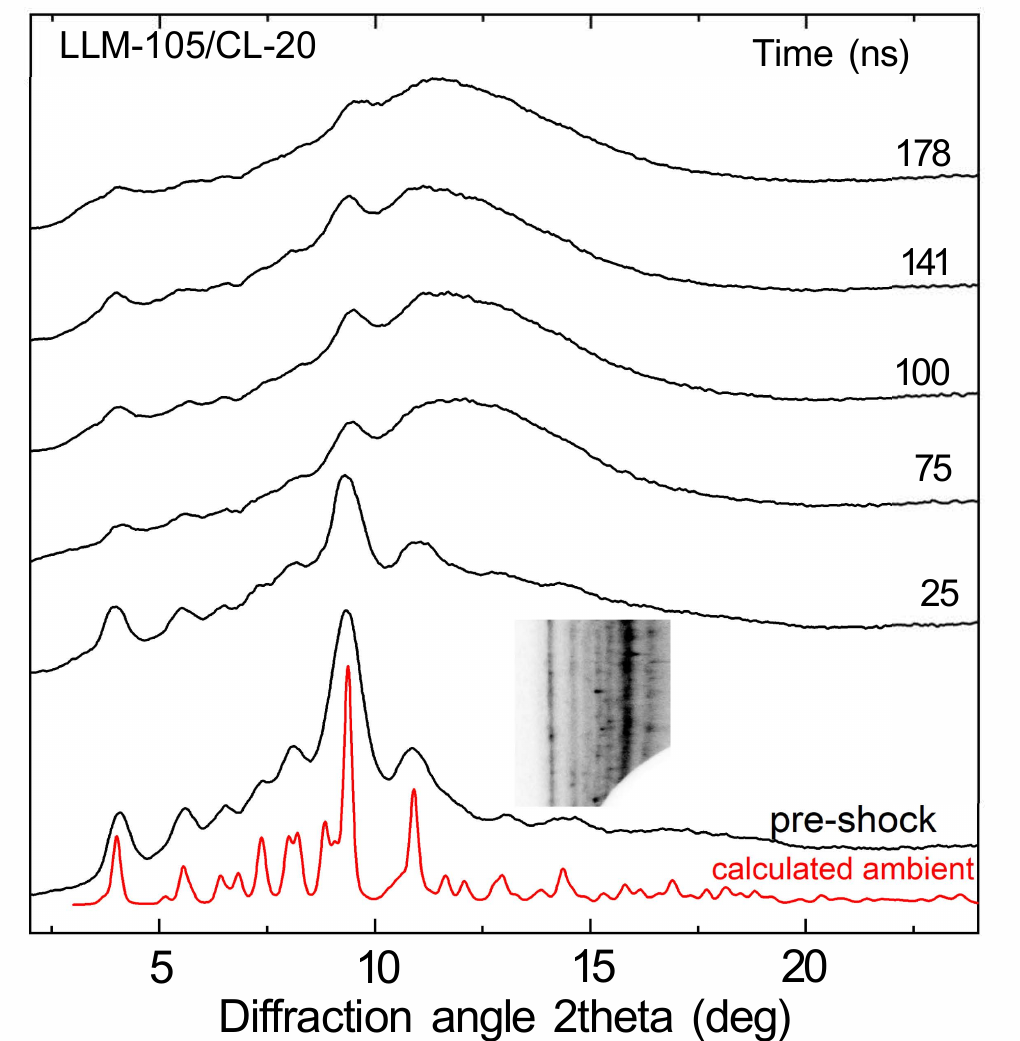}}
\centering
\caption{X-ray diffraction patterns of the LLM-105 PBX  under detonation induced, using a CL-20 based PBX, shock conditions at selected times. The calculated pattern of monoclinic LLM-105 \cite{Gilardi2001} at ambient conditions   is plotted with red.  Times are relative to the detonation front passing through the X-ray beam.  The inset 2D X-ray diffraction image  in rectangular coordinates (cake)  corresponds to the pre-shock pattern of the LLM-105 PBX.}
\end{figure}

In contrast to the case of the TATB-based specimens, no apparent shock/pressure induced shift of the corresponding Bragg peaks can be observed in the case of LLM-105. As clearly   seen in Fig. 5, at  times later than t=25 ns the patterns are indicative of a superposition of the most intense Bragg peaks of LLM-105 (presumably as a residual from earlier times) and a broad feature that signals the lack of long-range order.

\section{Discussion}
TATB, as opposed to LLM-105, exhibits diffraction behavior that is not consistent with prompt decomposition.  We consider various potential scenarios that could explain the observed TATB diffraction as well as the underlying structural behavior under detonation conditions, and the observed differences as compared to   LLM-105. The crystal structure of LLM-105 might not be as resilient under detonation-driven thermodynamic conditions, in contrast to TATB. This corroborates with the fact that TATB is more insensitive under shock conditions than LLM-105, although LLM-105 is also a candidate insensitive HE. The implications of the observed resilience  of TATB crystallites and associated implications are discussed further in the following paragraphs.

Moreover, we cannot discount the layered TATB structure that leads to an XRD pattern dominated by an intense 002 peak.  This high-intensity peak can be easily traced under pressure given that the intensity of the Bragg peaks at higher angles than the 002 have, at least, an order of magnitude lower intensity. In contrast, the XRD pattern of LLM-105 consists of closely spaced peaks of similar intensity. In other words, in the case of LLM-105 the position under high-pressure of a given Bragg peak might coincide with the 2$\theta$ position of the next Bragg peak at ambient conditions.  Due to the relatively large energy width of the incident X-ray beam, a definite conclusion about the validity of each scenario is difficult without excluding the possibility of a synergy ($i.e.$ higher resilience of TATB over LLM-105 together with experimental limitations) between the two scenarios. Additional measurements are needed, preferably with an almost monochromatic incident X-ray beam, to completely elucidate this issue. Moreover, a more planar detonation drive would reduce the time smearing caused by the shock front curvature in our experiments.


As mentioned in the results section, in the case of the TATB PBX only $\approx$30\% of crystalline TATB is observed under in a high-pressure state detonation-induced shock conditions. This could be attributed to a highly heterogenous   shock reactivity in the TATB PBX that  proceeds according to the \textquotedblleft hot spot\textquotedblright {}  mechanism \cite{Field1992,Campbell1961,Bowden1952}, where behind the leading edge of the shock, shrinking compressed crystalline TATB regions would coexist with growing regions of reaction products. This is what normally one would expect in a polycrystalline, see 2D images in Fig. 2(a), polymer bonded material.  As discussed in the experimental section, persistence in the scintillators of detector system may be largely responsible for the ambient=pressure peaks that persist, but we cannot discount the possibility that a small contribution to this ambient set of peaks arises at the cylinder edges, where a discontinuity in pressure exists, and the detonation may be releasing a small amount of undetonated material on these timescales.  (See also Figs. S1 and S3.)    It is noteworthy, that the evolution of SCTATB, in both conventional and colliding detonation setup,   under shock conditions  from single crystal to a multigrain morphology is also consistent with heterogenous reactivity.

Interestingly,and additionally,  the observation of the 30\% remaining TATB under shock conditions is in agreement with the basic principle  of the Zel'dovitch-von Neumann-Doering (ZND)  theory for detonation \cite{Kirkwood1954}.  In this context, the 30\% of the observed TATB under pressure  could represent the material at and immediately behind the thin von Neumann spike shock wave that compresses the explosive to a high pressure instantaneously, while the explosive remains unreacted. Given that in our experimental setup the incident X-rays are orthogonal to the curved shock front\cite{Bagge-Hansen2019,Hammons2019}, we do simultaneously probe a superposition of various states through the thin shock front and the reaction zone, but also note that we observe pressure decreasing as the TATB continues to decompose (Fig. 2) and observe essentially complete decomposition at very late times (Fig. 6). Thus, our study is consistent with the observation of decomposing TATB crystallites, sufficient to exhibit compressed 002 diffraction peak, that persist due to the  longer reaction zone of TATB relative to other explosives \cite{Tarver2006,Loboiko2000}.  The remaining 70\% of the TATB that does not contribute to the pressurized 002 peak is material that has decomposed and/or obliterated to  sufficiently small particulates that no longer produce meaningful diffraction.

Now we turn our attention to the broad feature observed in the patterns of both TATB and LLM-105 that signals the lack of long-range order. Although a detailed analysis on the origin and the characteristics (e.g. structural factor) of this feature is beyond the scope of this paper, this broad feature is attributed to XRD originating from non- crystalline states. The HE  undergoes a shock-induced chemical transformation under detonation-like conditions towards  non-crystalline detonation reaction products.  Fig. 6 shows the XRD patterns of the TATB  and LLM-105 PBX specimens at later times, beyond the expected reaction zone or  detonation-induced shock compression. For both the TATB and LLM-105 the 2$\theta$ position of the maximum of this feature shifts towards lower angles (higher $d$-spacings) with time. This is consistent with  the pressure release and the consequent increase of the interatomic/intermolecular distances with detonation product expansion. In the case of LLM-105, the absence of a crystalline XRD pattern at earlier times than TATB, under basically identical shock conditions, suggests a faster reactivity and is consistent with a shorter reaction zone in LLM-105.

\begin{figure}[ht]
{\includegraphics[width=\linewidth]{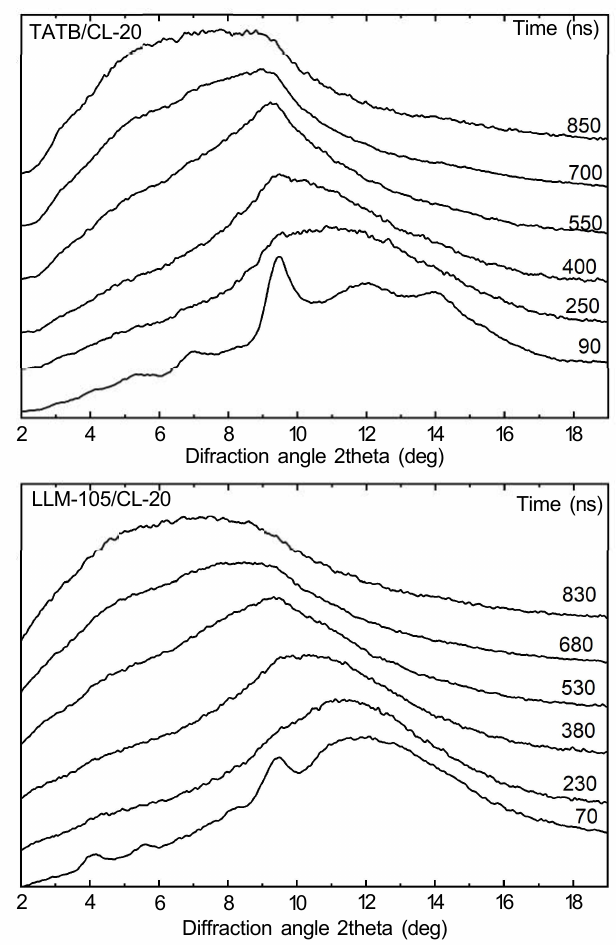}}
\centering
\caption{XRD patterns of TATB and LLM-105 under detonation induced, using a CL-20 based PBX, shock conditions at selected times.   Times are relative to the detonation front passing through the X-ray beam.  }
\end{figure}

Several theoretical and modeling papers are addressing TATB insensitivity; experimental capabilities are not yet able to address the time and length scales of the modeling and vice-versa.  In DFT and MD simulations of overdriven TATB crystals up to 430 ps, nitrogen heterocycle intermediates slow the TATB reaction particularly for the generation of ultimate N$_2$ and CO$_2$ products \cite{Manaa2009}.  The experiments herein are only able to resolve at 10's of ns timescales and measure TATB crystallinity, rather than directly probing the sub-ns chemistry.  Note, however, that if the heterocycles are forming in-plane, a 002-like out-of-plane diffraction peak could persist as these superstructures decompose into carbonaceous solids and product gases.  Recent theory demonstrates shear bands within the planar TATB crystallites under shock loading create localized reactive areas for TATB decomposition \cite{Kroonblawd2020}; these can be surrounded by TATB retaining its crystallinity.  Arrhenius fits to much shorter timescale quantum molecular dynamics show much of the TATB persisting over 100 ns timscales.  Again, although experiment cannot yet directly interrogate nanoscale shear banding, the experimental observation of macroscopically ~30\% TATB-like crystallinity under pressure and persisting over 100 ns timescales is consistent with the model extrapolation as well as the relatively long reaction zone in TATB.

\section{Conclusion}
The structural evolution of TATB was studied under detonation-induced shock conditions using in-situ synchrotron X-ray diffraction in the 100 ns timescale up to 60+ GPa. The results indicate that a portion of the TATB remains stable up to the highest observed  pressures attained during this study. Thus, our results support the ZND  theory for detonation.   Moreover, under moderate shock pressure ($<$35 GPa) the crystal structure of a portion of the  TATB also persists during pressure release. In contrast to the TATB, in the case of LLM-105 diffraction peaks under pressure during detonation were  not observed.  This can be attributed either to the vast insensitivity of TATB and/or to experimental limitations relevant to the energy width of the incident X-ray beam and the characteristics of the corresponding XRD patterns. The results indicate the surprising resilience of TATB under these high-pressure, temperature and shock conditions, providing a baseline for understanding the insensitivity of TATB.  The results also provide intriguing information for the extended reaction zone in TATB, and the hot-spot mechanisms for initiating and propagating detonation in this uniquely insensitive explosive. Our findings highlight differences between TATB and other explosives (LLM-105 presented herein) and give insight into the extreme insensitivity of TATB. TATB, under detonation conditions, exhibits diffraction peak shifting which may indicate a relatively unique response

\begin{acknowledgments}
This work was performed under the auspices of the U. S. Department of Energy by Lawrence Livermore National Security, LLC under Contract DE-AC52-07NA27344. We gratefully acknowledge the LLNL LDRD program for funding support of this project under 18-SI-004. We thank Ben Yancey for the  Laue measurements on the SC TATB samples. The Dynamic Compression Sector at the Advanced Photon Source is managed by Washington State University and funded by the National Nuclear Security Administration of the U.S. Department of Energy under Cooperative Agreement No. DE-NA0002442. Supporting experiments and data were also performed at 32-ID-B at APS. This research used resources of the Advanced Photon Source, a U.S. Department of Energy (DOE) Office of Science User Facility operated for the DOE Office of Science by Argonne National Laboratory under Contract No. DE-AC02–06CH11357.
\end{acknowledgments}


\begin{thebibliography}{49}%
\makeatletter
\providecommand \@ifxundefined [1]{%
 \@ifx{#1\undefined}
}%
\providecommand \@ifnum [1]{%
 \ifnum #1\expandafter \@firstoftwo
 \else \expandafter \@secondoftwo
 \fi
}%
\providecommand \@ifx [1]{%
 \ifx #1\expandafter \@firstoftwo
 \else \expandafter \@secondoftwo
 \fi
}%
\providecommand \natexlab [1]{#1}%
\providecommand \enquote  [1]{``#1''}%
\providecommand \bibnamefont  [1]{#1}%
\providecommand \bibfnamefont [1]{#1}%
\providecommand \citenamefont [1]{#1}%
\providecommand \href@noop [0]{\@secondoftwo}%
\providecommand \href [0]{\begingroup \@sanitize@url \@href}%
\providecommand \@href[1]{\@@startlink{#1}\@@href}%
\providecommand \@@href[1]{\endgroup#1\@@endlink}%
\providecommand \@sanitize@url [0]{\catcode `\\12\catcode `\$12\catcode
  `\&12\catcode `\#12\catcode `\^12\catcode `\_12\catcode `\%12\relax}%
\providecommand \@@startlink[1]{}%
\providecommand \@@endlink[0]{}%
\providecommand \url  [0]{\begingroup\@sanitize@url \@url }%
\providecommand \@url [1]{\endgroup\@href {#1}{\urlprefix }}%
\providecommand \urlprefix  [0]{URL }%
\providecommand \Eprint [0]{\href }%
\providecommand \doibase [0]{http://dx.doi.org/}%
\providecommand \selectlanguage [0]{\@gobble}%
\providecommand \bibinfo  [0]{\@secondoftwo}%
\providecommand \bibfield  [0]{\@secondoftwo}%
\providecommand \translation [1]{[#1]}%
\providecommand \BibitemOpen [0]{}%
\providecommand \bibitemStop [0]{}%
\providecommand \bibitemNoStop [0]{.\EOS\space}%
\providecommand \EOS [0]{\spacefactor3000\relax}%
\providecommand \BibitemShut  [1]{\csname bibitem#1\endcsname}%
\let\auto@bib@innerbib\@empty
\bibitem [{\citenamefont {Rice}\ and\ \citenamefont
  {Simpson}(1990)}]{Rice1990}%
  \BibitemOpen
  \bibfield  {author} {\bibinfo {author} {\bibfnamefont {S.}~\bibnamefont
  {Rice}}\ and\ \bibinfo {author} {\bibfnamefont {R.}~\bibnamefont {Simpson}},\
  }\href {\doibase 10.2172/6426268} {\emph {\bibinfo {title} {The unusual
  stability of {TATB} (1,3,5-triamino-2,4,6-trinitrobenzene): A review of the
  scientific literature}}},\ \bibinfo {type} {Tech. Rep.}\ (\bibinfo
  {institution} {Lawrence Livermore National Laboratory},\ \bibinfo {year}
  {1990})\BibitemShut {NoStop}%
\bibitem [{\citenamefont {Tarver}(2006)}]{Tarver2006}%
  \BibitemOpen
  \bibfield  {author} {\bibinfo {author} {\bibfnamefont {C.~M.}\ \bibnamefont
  {Tarver}},\ }\href {\doibase 10.1063/1.2263497} {\bibfield  {journal}
  {\bibinfo  {journal} {AIP Conference Proceedings}\ }\textbf {\bibinfo
  {volume} {845}},\ \bibinfo {pages} {1026} (\bibinfo {year}
  {2006})}\BibitemShut {NoStop}%
\bibitem [{\citenamefont {Manaa}\ and\ \citenamefont
  {Fried}(2012)}]{Manaa2012}%
  \BibitemOpen
  \bibfield  {author} {\bibinfo {author} {\bibfnamefont {M.~R.}\ \bibnamefont
  {Manaa}}\ and\ \bibinfo {author} {\bibfnamefont {L.~E.}\ \bibnamefont
  {Fried}},\ }\href {\doibase {10.1021/jp205920n}} {\bibfield  {journal}
  {\bibinfo  {journal} {J. Phys. Chem. C}\ }\textbf {\bibinfo {volume} {116}},\
  \bibinfo {pages} {2116} (\bibinfo {year} {2012})}\BibitemShut {NoStop}%
\bibitem [{\citenamefont {Rykounov}(2015)}]{Rykounov2015}%
  \BibitemOpen
  \bibfield  {author} {\bibinfo {author} {\bibfnamefont {A.~A.}\ \bibnamefont
  {Rykounov}},\ }\href@noop {} {\bibfield  {journal} {\bibinfo  {journal} {J.
  Appl. Phys.}\ }\textbf {\bibinfo {volume} {117}} (\bibinfo {year}
  {2015})}\BibitemShut {NoStop}%
\bibitem [{\citenamefont {Bedrov}\ \emph {et~al.}(2009)\citenamefont {Bedrov},
  \citenamefont {Borodin}, \citenamefont {Smith}, \citenamefont {Sewell},
  \citenamefont {Dattelbaum},\ and\ \citenamefont {Stevens}}]{Bedrov2009}%
  \BibitemOpen
  \bibfield  {author} {\bibinfo {author} {\bibfnamefont {D.}~\bibnamefont
  {Bedrov}}, \bibinfo {author} {\bibfnamefont {O.}~\bibnamefont {Borodin}},
  \bibinfo {author} {\bibfnamefont {G.~D.}\ \bibnamefont {Smith}}, \bibinfo
  {author} {\bibfnamefont {T.~D.}\ \bibnamefont {Sewell}}, \bibinfo {author}
  {\bibfnamefont {D.~M.}\ \bibnamefont {Dattelbaum}}, \ and\ \bibinfo {author}
  {\bibfnamefont {L.~L.}\ \bibnamefont {Stevens}},\ }\href@noop {} {\bibfield
  {journal} {\bibinfo  {journal} {J. Chem. Phys.}\ }\textbf {\bibinfo {volume}
  {131}} (\bibinfo {year} {2009})}\BibitemShut {NoStop}%
\bibitem [{\citenamefont {Byrd}\ and\ \citenamefont {Rice}(2007)}]{Byrd2007}%
  \BibitemOpen
  \bibfield  {author} {\bibinfo {author} {\bibfnamefont {E.~F.~C.}\
  \bibnamefont {Byrd}}\ and\ \bibinfo {author} {\bibfnamefont {B.~M.}\
  \bibnamefont {Rice}},\ }\href {\doibase 10.1021/jp0617930} {\bibfield
  {journal} {\bibinfo  {journal} {J. Phys. Chem. C}\ }\textbf {\bibinfo
  {volume} {111}},\ \bibinfo {pages} {2787} (\bibinfo {year}
  {2007})}\BibitemShut {NoStop}%
\bibitem [{\citenamefont {Wu}\ \emph {et~al.}(2014)\citenamefont {Wu},
  \citenamefont {Zhu},\ and\ \citenamefont {Xiao}}]{Wu2014}%
  \BibitemOpen
  \bibfield  {author} {\bibinfo {author} {\bibfnamefont {Q.}~\bibnamefont
  {Wu}}, \bibinfo {author} {\bibfnamefont {W.}~\bibnamefont {Zhu}}, \ and\
  \bibinfo {author} {\bibfnamefont {H.}~\bibnamefont {Xiao}},\ }\href {\doibase
  {10.1039/c4ra09123j}} {\bibfield  {journal} {\bibinfo  {journal} {RSC
  ADVANCES}\ }\textbf {\bibinfo {volume} {4}},\ \bibinfo {pages} {53149}
  (\bibinfo {year} {2014})}\BibitemShut {NoStop}%
\bibitem [{\citenamefont {Landerville}\ \emph {et~al.}(2010)\citenamefont
  {Landerville}, \citenamefont {Conroy}, \citenamefont {Budzevich},
  \citenamefont {Lin}, \citenamefont {White},\ and\ \citenamefont
  {Oleynik}}]{Landerville2010}%
  \BibitemOpen
  \bibfield  {author} {\bibinfo {author} {\bibfnamefont {A.~C.}\ \bibnamefont
  {Landerville}}, \bibinfo {author} {\bibfnamefont {M.~W.}\ \bibnamefont
  {Conroy}}, \bibinfo {author} {\bibfnamefont {M.~M.}\ \bibnamefont
  {Budzevich}}, \bibinfo {author} {\bibfnamefont {Y.}~\bibnamefont {Lin}},
  \bibinfo {author} {\bibfnamefont {C.~T.}\ \bibnamefont {White}}, \ and\
  \bibinfo {author} {\bibfnamefont {I.~I.}\ \bibnamefont {Oleynik}},\
  }\href@noop {} {\bibfield  {journal} {\bibinfo  {journal} {Appl. Phys.
  Lett.}\ }\textbf {\bibinfo {volume} {97}} (\bibinfo {year}
  {2010})}\BibitemShut {NoStop}%
\bibitem [{\citenamefont {Qin}\ \emph {et~al.}(2019)\citenamefont {Qin},
  \citenamefont {Yan}, \citenamefont {Zhong}, \citenamefont {Jiang},
  \citenamefont {Liu}, \citenamefont {Tang},\ and\ \citenamefont
  {Liu}}]{Qin2019}%
  \BibitemOpen
  \bibfield  {author} {\bibinfo {author} {\bibfnamefont {H.}~\bibnamefont
  {Qin}}, \bibinfo {author} {\bibfnamefont {B.-L.}\ \bibnamefont {Yan}},
  \bibinfo {author} {\bibfnamefont {M.}~\bibnamefont {Zhong}}, \bibinfo
  {author} {\bibfnamefont {C.-L.}\ \bibnamefont {Jiang}}, \bibinfo {author}
  {\bibfnamefont {F.-S.}\ \bibnamefont {Liu}}, \bibinfo {author} {\bibfnamefont
  {B.}~\bibnamefont {Tang}}, \ and\ \bibinfo {author} {\bibfnamefont {Q.-J.}\
  \bibnamefont {Liu}},\ }\href {\doibase {10.1016/j.physb.2018.10.003}}
  {\bibfield  {journal} {\bibinfo  {journal} {Physica B condens.}\ }\textbf
  {\bibinfo {volume} {552}},\ \bibinfo {pages} {151} (\bibinfo {year}
  {2019})}\BibitemShut {NoStop}%
\bibitem [{\citenamefont {Budzevich}\ \emph {et~al.}(2010)\citenamefont
  {Budzevich}, \citenamefont {Landerville}, \citenamefont {Conroy},
  \citenamefont {Lin}, \citenamefont {Oleynik},\ and\ \citenamefont
  {White}}]{Budzevich2010}%
  \BibitemOpen
  \bibfield  {author} {\bibinfo {author} {\bibfnamefont {M.~M.}\ \bibnamefont
  {Budzevich}}, \bibinfo {author} {\bibfnamefont {A.~C.}\ \bibnamefont
  {Landerville}}, \bibinfo {author} {\bibfnamefont {M.~W.}\ \bibnamefont
  {Conroy}}, \bibinfo {author} {\bibfnamefont {Y.}~\bibnamefont {Lin}},
  \bibinfo {author} {\bibfnamefont {I.~I.}\ \bibnamefont {Oleynik}}, \ and\
  \bibinfo {author} {\bibfnamefont {C.~T.}\ \bibnamefont {White}},\ }\href@noop
  {} {\bibfield  {journal} {\bibinfo  {journal} {J. Appl. Phys.}\ }\textbf
  {\bibinfo {volume} {107}} (\bibinfo {year} {2010})}\BibitemShut {NoStop}%
\bibitem [{\citenamefont {Fedorov}\ and\ \citenamefont
  {Zhuravlev}(2014)}]{Fedorov2014}%
  \BibitemOpen
  \bibfield  {author} {\bibinfo {author} {\bibfnamefont {I.~A.}\ \bibnamefont
  {Fedorov}}\ and\ \bibinfo {author} {\bibfnamefont {Y.~N.}\ \bibnamefont
  {Zhuravlev}},\ }\href {\doibase {10.1016/j.chemphys.2014.03.013}} {\bibfield
  {journal} {\bibinfo  {journal} {Chem. Phys.}\ }\textbf {\bibinfo {volume}
  {436}},\ \bibinfo {pages} {1} (\bibinfo {year} {2014})}\BibitemShut {NoStop}%
\bibitem [{\citenamefont {Kroonblawd}\ and\ \citenamefont
  {Fried}(2020)}]{Kroonblawd2020}%
  \BibitemOpen
  \bibfield  {author} {\bibinfo {author} {\bibfnamefont {M.~P.}\ \bibnamefont
  {Kroonblawd}}\ and\ \bibinfo {author} {\bibfnamefont {L.~E.}\ \bibnamefont
  {Fried}},\ }\href {\doibase 10.1103/PhysRevLett.124.206002} {\bibfield
  {journal} {\bibinfo  {journal} {Phys. Rev. Lett.}\ }\textbf {\bibinfo
  {volume} {124}},\ \bibinfo {pages} {206002} (\bibinfo {year}
  {2020})}\BibitemShut {NoStop}%
\bibitem [{\citenamefont {Stevens}\ \emph {et~al.}(2008)\citenamefont
  {Stevens}, \citenamefont {Velisavljevic}, \citenamefont {Hooks},\ and\
  \citenamefont {Dattelbaum}}]{Stevens2008}%
  \BibitemOpen
  \bibfield  {author} {\bibinfo {author} {\bibfnamefont {L.~L.}\ \bibnamefont
  {Stevens}}, \bibinfo {author} {\bibfnamefont {N.}~\bibnamefont
  {Velisavljevic}}, \bibinfo {author} {\bibfnamefont {D.~E.}\ \bibnamefont
  {Hooks}}, \ and\ \bibinfo {author} {\bibfnamefont {D.~M.}\ \bibnamefont
  {Dattelbaum}},\ }\href@noop {} {\bibfield  {journal} {\bibinfo  {journal}
  {Propellants Explos. Pyrotech.}\ }\textbf {\bibinfo {volume} {33}},\ \bibinfo
  {pages} {286} (\bibinfo {year} {2008})}\BibitemShut {NoStop}%
\bibitem [{\citenamefont {Plisson}\ \emph {et~al.}(2017)\citenamefont
  {Plisson}, \citenamefont {Pineau}, \citenamefont {Weck}, \citenamefont
  {Bruneton}, \citenamefont {Guignot},\ and\ \citenamefont
  {Loubeyre}}]{Plisson2017}%
  \BibitemOpen
  \bibfield  {author} {\bibinfo {author} {\bibfnamefont {T.}~\bibnamefont
  {Plisson}}, \bibinfo {author} {\bibfnamefont {N.}~\bibnamefont {Pineau}},
  \bibinfo {author} {\bibfnamefont {G.}~\bibnamefont {Weck}}, \bibinfo {author}
  {\bibfnamefont {E.}~\bibnamefont {Bruneton}}, \bibinfo {author}
  {\bibfnamefont {N.}~\bibnamefont {Guignot}}, \ and\ \bibinfo {author}
  {\bibfnamefont {P.}~\bibnamefont {Loubeyre}},\ }\href@noop {} {\bibfield
  {journal} {\bibinfo  {journal} {J. Appl. Phys.}\ }\textbf {\bibinfo {volume}
  {122}},\ \bibinfo {pages} {235901} (\bibinfo {year} {2017})}\BibitemShut
  {NoStop}%
\bibitem [{\citenamefont {Steele}\ \emph {et~al.}(2019)\citenamefont {Steele},
  \citenamefont {Clarke}, \citenamefont {Kroonblawd}, \citenamefont {Kuo},
  \citenamefont {Pagoria}, \citenamefont {Tkachev}, \citenamefont {Smith},
  \citenamefont {Bastea}, \citenamefont {Fried}, \citenamefont {Zaug},
  \citenamefont {Stavrou},\ and\ \citenamefont {Tschauner}}]{Steele2019a}%
  \BibitemOpen
  \bibfield  {author} {\bibinfo {author} {\bibfnamefont {B.~A.}\ \bibnamefont
  {Steele}}, \bibinfo {author} {\bibfnamefont {S.~M.}\ \bibnamefont {Clarke}},
  \bibinfo {author} {\bibfnamefont {M.~P.}\ \bibnamefont {Kroonblawd}},
  \bibinfo {author} {\bibfnamefont {I.-F.~W.}\ \bibnamefont {Kuo}}, \bibinfo
  {author} {\bibfnamefont {P.~F.}\ \bibnamefont {Pagoria}}, \bibinfo {author}
  {\bibfnamefont {S.~N.}\ \bibnamefont {Tkachev}}, \bibinfo {author}
  {\bibfnamefont {J.~S.}\ \bibnamefont {Smith}}, \bibinfo {author}
  {\bibfnamefont {S.}~\bibnamefont {Bastea}}, \bibinfo {author} {\bibfnamefont
  {L.~E.}\ \bibnamefont {Fried}}, \bibinfo {author} {\bibfnamefont {J.~M.}\
  \bibnamefont {Zaug}}, \bibinfo {author} {\bibfnamefont {E.}~\bibnamefont
  {Stavrou}}, \ and\ \bibinfo {author} {\bibfnamefont {O.}~\bibnamefont
  {Tschauner}},\ }\href {\doibase 10.1063/1.5091947} {\bibfield  {journal}
  {\bibinfo  {journal} {Applied Physics Letters}\ }\textbf {\bibinfo {volume}
  {114}},\ \bibinfo {pages} {191901} (\bibinfo {year} {2019})}\BibitemShut
  {NoStop}%
\bibitem [{\citenamefont {Gustavsen}\ \emph {et~al.}(2006)\citenamefont
  {Gustavsen}, \citenamefont {Sheffield},\ and\ \citenamefont
  {Alcon}}]{Gustavsen2006}%
  \BibitemOpen
  \bibfield  {author} {\bibinfo {author} {\bibfnamefont {R.~L.}\ \bibnamefont
  {Gustavsen}}, \bibinfo {author} {\bibfnamefont {S.~A.}\ \bibnamefont
  {Sheffield}}, \ and\ \bibinfo {author} {\bibfnamefont {R.~R.}\ \bibnamefont
  {Alcon}},\ }\href@noop {} {\bibfield  {journal} {\bibinfo  {journal} {J.
  Appl. Phys.}\ }\textbf {\bibinfo {volume} {99}},\ \bibinfo {pages} {114907}
  (\bibinfo {year} {2006})}\BibitemShut {NoStop}%
\bibitem [{\citenamefont {Dick}\ \emph {et~al.}(1988)\citenamefont {Dick},
  \citenamefont {Forest}, \citenamefont {Ramsay},\ and\ \citenamefont
  {Seitz}}]{Dick1988}%
  \BibitemOpen
  \bibfield  {author} {\bibinfo {author} {\bibfnamefont {J.~J.}\ \bibnamefont
  {Dick}}, \bibinfo {author} {\bibfnamefont {C.~A.}\ \bibnamefont {Forest}},
  \bibinfo {author} {\bibfnamefont {J.~B.}\ \bibnamefont {Ramsay}}, \ and\
  \bibinfo {author} {\bibfnamefont {W.~L.}\ \bibnamefont {Seitz}},\ }\href@noop
  {} {\bibfield  {journal} {\bibinfo  {journal} {J. Appl. Phys.}\ }\textbf
  {\bibinfo {volume} {63}},\ \bibinfo {pages} {4881} (\bibinfo {year}
  {1988})}\BibitemShut {NoStop}%
\bibitem [{\citenamefont {Jackson}\ \emph {et~al.}(1976)\citenamefont
  {Jackson}, \citenamefont {Green}, \citenamefont {Barlett}, \citenamefont
  {Hofer}, \citenamefont {Kramer}, \citenamefont {Lee}, \citenamefont
  {EJ.~Nidick}, \citenamefont {Shaw},\ and\ \citenamefont
  {Weingart}}]{Jackson1976}%
  \BibitemOpen
  \bibfield  {author} {\bibinfo {author} {\bibfnamefont {R.~K.}\ \bibnamefont
  {Jackson}}, \bibinfo {author} {\bibfnamefont {L.~G.}\ \bibnamefont {Green}},
  \bibinfo {author} {\bibfnamefont {R.~H.}\ \bibnamefont {Barlett}}, \bibinfo
  {author} {\bibfnamefont {W.~C.~D.}\ \bibnamefont {Hofer}}, \bibinfo {author}
  {\bibfnamefont {P.~E.}\ \bibnamefont {Kramer}}, \bibinfo {author}
  {\bibfnamefont {R.~S.}\ \bibnamefont {Lee}}, \bibinfo {author} {\bibfnamefont
  {J.}~\bibnamefont {EJ.~Nidick}}, \bibinfo {author} {\bibfnamefont {L.~L.}\
  \bibnamefont {Shaw}}, \ and\ \bibinfo {author} {\bibfnamefont {R.~C.}\
  \bibnamefont {Weingart}},\ }in\ \href@noop {} {\emph {\bibinfo {booktitle}
  {Sixth Symposium (International) on Detonation}}},\ Vol.\ \bibinfo {volume}
  {ONR ACR-221}\ (\bibinfo {year} {1976})\ p.\ \bibinfo {pages}
  {755}\BibitemShut {NoStop}%
\bibitem [{\citenamefont {Dallman}\ and\ \citenamefont
  {Wackerle}(1993)}]{Dallman1993}%
  \BibitemOpen
  \bibfield  {author} {\bibinfo {author} {\bibfnamefont {J.}~\bibnamefont
  {Dallman}}\ and\ \bibinfo {author} {\bibfnamefont {J.}~\bibnamefont
  {Wackerle}},\ }in\ \href@noop {} {\emph {\bibinfo {booktitle} {Tenth
  Symposium (International) on Detonation}}},\ Vol.\ \bibinfo {volume} {ONR
  33395-12},\ \bibinfo {editor} {edited by\ \bibinfo {editor} {\bibfnamefont
  {J.~M.~S.}\ \bibnamefont {{(Office of Naval Research, Arlington, VA)}}}}\
  (\bibinfo {year} {1993})\ p.\ \bibinfo {pages} {130}\BibitemShut {NoStop}%
\bibitem [{\citenamefont {Marshall}\ \emph {et~al.}(2020)\citenamefont
  {Marshall}, \citenamefont {Fernandez-Pañella}, \citenamefont {Myers},
  \citenamefont {Eggert}, \citenamefont {Erskine}, \citenamefont {Bastea},
  \citenamefont {Fried},\ and\ \citenamefont {Leininger}}]{Marshall2020}%
  \BibitemOpen
  \bibfield  {author} {\bibinfo {author} {\bibfnamefont {M.~C.}\ \bibnamefont
  {Marshall}}, \bibinfo {author} {\bibfnamefont {A.}~\bibnamefont
  {Fernandez-Pañella}}, \bibinfo {author} {\bibfnamefont {T.~W.}\ \bibnamefont
  {Myers}}, \bibinfo {author} {\bibfnamefont {J.~H.}\ \bibnamefont {Eggert}},
  \bibinfo {author} {\bibfnamefont {D.~J.}\ \bibnamefont {Erskine}}, \bibinfo
  {author} {\bibfnamefont {S.}~\bibnamefont {Bastea}}, \bibinfo {author}
  {\bibfnamefont {L.~E.}\ \bibnamefont {Fried}}, \ and\ \bibinfo {author}
  {\bibfnamefont {L.~D.}\ \bibnamefont {Leininger}},\ }\href {\doibase
  10.1063/5.0005818} {\bibfield  {journal} {\bibinfo  {journal} {Journal of
  Applied Physics}\ }\textbf {\bibinfo {volume} {127}},\ \bibinfo {pages}
  {185901} (\bibinfo {year} {2020})}\BibitemShut {NoStop}%
\bibitem [{\citenamefont {Nandi}\ \emph {et~al.}(2007)\citenamefont {Nandi},
  \citenamefont {Kasar}, \citenamefont {Thanigaivelan}, \citenamefont {Ghosh},
  \citenamefont {Mandal},\ and\ \citenamefont {Bhattacharyya}}]{Nandi2007}%
  \BibitemOpen
  \bibfield  {author} {\bibinfo {author} {\bibfnamefont {A.~K.}\ \bibnamefont
  {Nandi}}, \bibinfo {author} {\bibfnamefont {S.~M.}\ \bibnamefont {Kasar}},
  \bibinfo {author} {\bibfnamefont {U.}~\bibnamefont {Thanigaivelan}}, \bibinfo
  {author} {\bibfnamefont {M.}~\bibnamefont {Ghosh}}, \bibinfo {author}
  {\bibfnamefont {A.~K.}\ \bibnamefont {Mandal}}, \ and\ \bibinfo {author}
  {\bibfnamefont {S.~C.}\ \bibnamefont {Bhattacharyya}},\ }\href {\doibase
  10.1080/07370650701567066} {\bibfield  {journal} {\bibinfo  {journal}
  {Journal of Energetic Materials}\ }\textbf {\bibinfo {volume} {25}},\
  \bibinfo {pages} {213} (\bibinfo {year} {2007})}\BibitemShut {NoStop}%
\bibitem [{\citenamefont {Foltz}\ \emph {et~al.}(1996)\citenamefont {Foltz},
  \citenamefont {Maienschein},\ and\ \citenamefont {Green}}]{Foltz1996}%
  \BibitemOpen
  \bibfield  {author} {\bibinfo {author} {\bibfnamefont {M.~F.}\ \bibnamefont
  {Foltz}}, \bibinfo {author} {\bibfnamefont {J.~L.}\ \bibnamefont
  {Maienschein}}, \ and\ \bibinfo {author} {\bibfnamefont {L.~G.}\ \bibnamefont
  {Green}},\ }\href {https://doi.org/10.1007/BF00372187} {\bibfield  {journal}
  {\bibinfo  {journal} {Journal of Materials Science}\ }\textbf {\bibinfo
  {volume} {31}},\ \bibinfo {pages} {1741} (\bibinfo {year}
  {1996})}\BibitemShut {NoStop}%
\bibitem [{\citenamefont {Han}\ \emph {et~al.}(2009)\citenamefont {Han},
  \citenamefont {Pagoria}, \citenamefont {Gash}, \citenamefont {Maiti},
  \citenamefont {Orme}, \citenamefont {Mitchell},\ and\ \citenamefont
  {Fried}}]{Han2009}%
  \BibitemOpen
  \bibfield  {author} {\bibinfo {author} {\bibfnamefont {T.~Y.-J.}\
  \bibnamefont {Han}}, \bibinfo {author} {\bibfnamefont {P.~F.}\ \bibnamefont
  {Pagoria}}, \bibinfo {author} {\bibfnamefont {A.~E.}\ \bibnamefont {Gash}},
  \bibinfo {author} {\bibfnamefont {A.}~\bibnamefont {Maiti}}, \bibinfo
  {author} {\bibfnamefont {C.~A.}\ \bibnamefont {Orme}}, \bibinfo {author}
  {\bibfnamefont {A.~R.}\ \bibnamefont {Mitchell}}, \ and\ \bibinfo {author}
  {\bibfnamefont {L.~E.}\ \bibnamefont {Fried}},\ }\href {\doibase
  10.1039/B810109D} {\bibfield  {journal} {\bibinfo  {journal} {New J. Chem.}\
  }\textbf {\bibinfo {volume} {33}},\ \bibinfo {pages} {50} (\bibinfo {year}
  {2009})}\BibitemShut {NoStop}%
\bibitem [{\citenamefont {Peterson}\ and\ \citenamefont
  {Idar}(2005)}]{Peterson2005}%
  \BibitemOpen
  \bibfield  {author} {\bibinfo {author} {\bibfnamefont {P.}~\bibnamefont
  {Peterson}}\ and\ \bibinfo {author} {\bibfnamefont {D.}~\bibnamefont
  {Idar}},\ }\href {\doibase https://doi.org/10.1002/prep.200400088} {\bibfield
   {journal} {\bibinfo  {journal} {Propellants, Explosives, Pyrotechnics}\
  }\textbf {\bibinfo {volume} {30}},\ \bibinfo {pages} {88} (\bibinfo {year}
  {2005})}\BibitemShut {NoStop}%
\bibitem [{\citenamefont {Bagge-Hansen}\ \emph {et~al.}(2015)\citenamefont
  {Bagge-Hansen}, \citenamefont {Lauderbach}, \citenamefont {Hodgin},
  \citenamefont {Bastea}, \citenamefont {Fried}, \citenamefont {Jones},
  \citenamefont {van Buuren}, \citenamefont {Hansen}, \citenamefont {Benterou},
  \citenamefont {May}, \citenamefont {Graber}, \citenamefont {Jensen},
  \citenamefont {Ilavsky},\ and\ \citenamefont {Willey}}]{Bagge-Hansen2015}%
  \BibitemOpen
  \bibfield  {author} {\bibinfo {author} {\bibfnamefont {M.}~\bibnamefont
  {Bagge-Hansen}}, \bibinfo {author} {\bibfnamefont {L.}~\bibnamefont
  {Lauderbach}}, \bibinfo {author} {\bibfnamefont {R.}~\bibnamefont {Hodgin}},
  \bibinfo {author} {\bibfnamefont {S.}~\bibnamefont {Bastea}}, \bibinfo
  {author} {\bibfnamefont {L.}~\bibnamefont {Fried}}, \bibinfo {author}
  {\bibfnamefont {A.}~\bibnamefont {Jones}}, \bibinfo {author} {\bibfnamefont
  {T.}~\bibnamefont {van Buuren}}, \bibinfo {author} {\bibfnamefont
  {D.}~\bibnamefont {Hansen}}, \bibinfo {author} {\bibfnamefont
  {J.}~\bibnamefont {Benterou}}, \bibinfo {author} {\bibfnamefont
  {C.}~\bibnamefont {May}}, \bibinfo {author} {\bibfnamefont {T.}~\bibnamefont
  {Graber}}, \bibinfo {author} {\bibfnamefont {B.~J.}\ \bibnamefont {Jensen}},
  \bibinfo {author} {\bibfnamefont {J.}~\bibnamefont {Ilavsky}}, \ and\
  \bibinfo {author} {\bibfnamefont {T.~M.}\ \bibnamefont {Willey}},\ }\href
  {\doibase 10.1063/1.4922866} {\bibfield  {journal} {\bibinfo  {journal} {J.
  Appl. Phys.}\ }\textbf {\bibinfo {volume} {117}},\ \bibinfo {pages} {245902}
  (\bibinfo {year} {2015})}\BibitemShut {NoStop}%
\bibitem [{\citenamefont {Watkins}\ \emph {et~al.}(2017)\citenamefont
  {Watkins}, \citenamefont {Velizhanin}, \citenamefont {Dattelbaum},
  \citenamefont {Gustavsen}, \citenamefont {Aslam}, \citenamefont {Podlesak},
  \citenamefont {Huber}, \citenamefont {Firestone}, \citenamefont {Ringstrand},
  \citenamefont {Willey}, \citenamefont {Bagge-Hansen}, \citenamefont {Hodgin},
  \citenamefont {Lauderbach}, \citenamefont {van Buuren}, \citenamefont
  {Sinclair}, \citenamefont {Rigg}, \citenamefont {Seifert},\ and\
  \citenamefont {Gog}}]{Watkins2017}%
  \BibitemOpen
  \bibfield  {author} {\bibinfo {author} {\bibfnamefont {E.~B.}\ \bibnamefont
  {Watkins}}, \bibinfo {author} {\bibfnamefont {K.~A.}\ \bibnamefont
  {Velizhanin}}, \bibinfo {author} {\bibfnamefont {D.~M.}\ \bibnamefont
  {Dattelbaum}}, \bibinfo {author} {\bibfnamefont {R.~L.}\ \bibnamefont
  {Gustavsen}}, \bibinfo {author} {\bibfnamefont {T.~D.}\ \bibnamefont
  {Aslam}}, \bibinfo {author} {\bibfnamefont {D.~W.}\ \bibnamefont {Podlesak}},
  \bibinfo {author} {\bibfnamefont {R.~C.}\ \bibnamefont {Huber}}, \bibinfo
  {author} {\bibfnamefont {M.~A.}\ \bibnamefont {Firestone}}, \bibinfo {author}
  {\bibfnamefont {B.~S.}\ \bibnamefont {Ringstrand}}, \bibinfo {author}
  {\bibfnamefont {T.~M.}\ \bibnamefont {Willey}}, \bibinfo {author}
  {\bibfnamefont {M.}~\bibnamefont {Bagge-Hansen}}, \bibinfo {author}
  {\bibfnamefont {R.}~\bibnamefont {Hodgin}}, \bibinfo {author} {\bibfnamefont
  {L.}~\bibnamefont {Lauderbach}}, \bibinfo {author} {\bibfnamefont
  {T.}~\bibnamefont {van Buuren}}, \bibinfo {author} {\bibfnamefont
  {N.}~\bibnamefont {Sinclair}}, \bibinfo {author} {\bibfnamefont {P.~A.}\
  \bibnamefont {Rigg}}, \bibinfo {author} {\bibfnamefont {S.}~\bibnamefont
  {Seifert}}, \ and\ \bibinfo {author} {\bibfnamefont {T.}~\bibnamefont
  {Gog}},\ }\href {\doibase 10.1021/acs.jpcc.7b05637} {\bibfield  {journal}
  {\bibinfo  {journal} {J. Phys. Chem. C}\ }\textbf {\bibinfo {volume} {121}},\
  \bibinfo {pages} {23129} (\bibinfo {year} {2017})}\BibitemShut {NoStop}%
\bibitem [{\citenamefont {Willey}\ \emph {et~al.}(2016)\citenamefont {Willey},
  \citenamefont {Champley}, \citenamefont {Hodgin}, \citenamefont {Lauderbach},
  \citenamefont {Bagge-Hansen}, \citenamefont {May}, \citenamefont {Sanchez},
  \citenamefont {Jensen}, \citenamefont {Iverson},\ and\ \citenamefont {van
  Buuren}}]{Willey2016}%
  \BibitemOpen
  \bibfield  {author} {\bibinfo {author} {\bibfnamefont {T.~M.}\ \bibnamefont
  {Willey}}, \bibinfo {author} {\bibfnamefont {K.}~\bibnamefont {Champley}},
  \bibinfo {author} {\bibfnamefont {R.}~\bibnamefont {Hodgin}}, \bibinfo
  {author} {\bibfnamefont {L.}~\bibnamefont {Lauderbach}}, \bibinfo {author}
  {\bibfnamefont {M.}~\bibnamefont {Bagge-Hansen}}, \bibinfo {author}
  {\bibfnamefont {C.}~\bibnamefont {May}}, \bibinfo {author} {\bibfnamefont
  {N.}~\bibnamefont {Sanchez}}, \bibinfo {author} {\bibfnamefont {B.~J.}\
  \bibnamefont {Jensen}}, \bibinfo {author} {\bibfnamefont {A.}~\bibnamefont
  {Iverson}}, \ and\ \bibinfo {author} {\bibfnamefont {T.}~\bibnamefont {van
  Buuren}},\ }\href {\doibase 10.1063/1.4953681} {\bibfield  {journal}
  {\bibinfo  {journal} {J. Appl. Phys.}\ }\textbf {\bibinfo {volume} {119}},\
  \bibinfo {pages} {235901} (\bibinfo {year} {2016})}\BibitemShut {NoStop}%
\bibitem [{\citenamefont {Bagge-Hansen}\ \emph {et~al.}(2019)\citenamefont
  {Bagge-Hansen}, \citenamefont {Bastea}, \citenamefont {Hammons},
  \citenamefont {Nielsen}, \citenamefont {Lauderbach}, \citenamefont {Hodgin},
  \citenamefont {Pagoria}, \citenamefont {May}, \citenamefont {Aloni},
  \citenamefont {Jones}, \citenamefont {Shaw}, \citenamefont {Bukovsky},
  \citenamefont {Sinclair}, \citenamefont {Gustavsen}, \citenamefont {Watkins},
  \citenamefont {Jensen}, \citenamefont {Dattelbaum}, \citenamefont
  {Firestone}, \citenamefont {Huber}, \citenamefont {Ringstrand}, \citenamefont
  {Lee}, \citenamefont {van Buuren}, \citenamefont {Fried},\ and\ \citenamefont
  {Willey}}]{Bagge-Hansen2019}%
  \BibitemOpen
  \bibfield  {author} {\bibinfo {author} {\bibfnamefont {M.}~\bibnamefont
  {Bagge-Hansen}}, \bibinfo {author} {\bibfnamefont {S.}~\bibnamefont
  {Bastea}}, \bibinfo {author} {\bibfnamefont {J.~A.}\ \bibnamefont {Hammons}},
  \bibinfo {author} {\bibfnamefont {M.~H.}\ \bibnamefont {Nielsen}}, \bibinfo
  {author} {\bibfnamefont {L.~M.}\ \bibnamefont {Lauderbach}}, \bibinfo
  {author} {\bibfnamefont {R.~L.}\ \bibnamefont {Hodgin}}, \bibinfo {author}
  {\bibfnamefont {P.}~\bibnamefont {Pagoria}}, \bibinfo {author} {\bibfnamefont
  {C.}~\bibnamefont {May}}, \bibinfo {author} {\bibfnamefont {S.}~\bibnamefont
  {Aloni}}, \bibinfo {author} {\bibfnamefont {A.}~\bibnamefont {Jones}},
  \bibinfo {author} {\bibfnamefont {W.~L.}\ \bibnamefont {Shaw}}, \bibinfo
  {author} {\bibfnamefont {E.~V.}\ \bibnamefont {Bukovsky}}, \bibinfo {author}
  {\bibfnamefont {N.}~\bibnamefont {Sinclair}}, \bibinfo {author}
  {\bibfnamefont {R.~L.}\ \bibnamefont {Gustavsen}}, \bibinfo {author}
  {\bibfnamefont {E.~B.}\ \bibnamefont {Watkins}}, \bibinfo {author}
  {\bibfnamefont {B.~J.}\ \bibnamefont {Jensen}}, \bibinfo {author}
  {\bibfnamefont {D.~M.}\ \bibnamefont {Dattelbaum}}, \bibinfo {author}
  {\bibfnamefont {M.~A.}\ \bibnamefont {Firestone}}, \bibinfo {author}
  {\bibfnamefont {R.~C.}\ \bibnamefont {Huber}}, \bibinfo {author}
  {\bibfnamefont {B.~S.}\ \bibnamefont {Ringstrand}}, \bibinfo {author}
  {\bibfnamefont {J.~R.~I.}\ \bibnamefont {Lee}}, \bibinfo {author}
  {\bibfnamefont {T.}~\bibnamefont {van Buuren}}, \bibinfo {author}
  {\bibfnamefont {L.~E.}\ \bibnamefont {Fried}}, \ and\ \bibinfo {author}
  {\bibfnamefont {T.~M.}\ \bibnamefont {Willey}},\ }\href
  {https://doi.org/10.1038/s41467-019-11666-z} {\bibfield  {journal} {\bibinfo
  {journal} {Nat. Commun.}\ }\textbf {\bibinfo {volume} {10}},\ \bibinfo
  {pages} {3819} (\bibinfo {year} {2019})}\BibitemShut {NoStop}%
\bibitem [{\citenamefont {Hammons}\ \emph {et~al.}(2019)\citenamefont
  {Hammons}, \citenamefont {Nielsen}, \citenamefont {Bagge-Hansen},
  \citenamefont {Bastea}, \citenamefont {Shaw}, \citenamefont {Lee},
  \citenamefont {Ilavsky}, \citenamefont {Sinclair}, \citenamefont {Fezzaa},
  \citenamefont {Lauderbach}, \citenamefont {Hodgin}, \citenamefont
  {Orlikowski}, \citenamefont {Fried},\ and\ \citenamefont
  {Willey}}]{Hammons2019}%
  \BibitemOpen
  \bibfield  {author} {\bibinfo {author} {\bibfnamefont {J.~A.}\ \bibnamefont
  {Hammons}}, \bibinfo {author} {\bibfnamefont {M.~H.}\ \bibnamefont
  {Nielsen}}, \bibinfo {author} {\bibfnamefont {M.}~\bibnamefont
  {Bagge-Hansen}}, \bibinfo {author} {\bibfnamefont {S.}~\bibnamefont
  {Bastea}}, \bibinfo {author} {\bibfnamefont {W.~L.}\ \bibnamefont {Shaw}},
  \bibinfo {author} {\bibfnamefont {J.~R.~I.}\ \bibnamefont {Lee}}, \bibinfo
  {author} {\bibfnamefont {J.}~\bibnamefont {Ilavsky}}, \bibinfo {author}
  {\bibfnamefont {N.}~\bibnamefont {Sinclair}}, \bibinfo {author}
  {\bibfnamefont {K.}~\bibnamefont {Fezzaa}}, \bibinfo {author} {\bibfnamefont
  {L.~M.}\ \bibnamefont {Lauderbach}}, \bibinfo {author} {\bibfnamefont
  {R.~L.}\ \bibnamefont {Hodgin}}, \bibinfo {author} {\bibfnamefont {D.~A.}\
  \bibnamefont {Orlikowski}}, \bibinfo {author} {\bibfnamefont {L.~E.}\
  \bibnamefont {Fried}}, \ and\ \bibinfo {author} {\bibfnamefont {T.~M.}\
  \bibnamefont {Willey}},\ }\href {\doibase 10.1021/acs.jpcc.9b02692}
  {\bibfield  {journal} {\bibinfo  {journal} {J. Phys. Chem. C}\ }\textbf
  {\bibinfo {volume} {123}},\ \bibinfo {pages} {19153} (\bibinfo {year}
  {2019})}\BibitemShut {NoStop}%
\bibitem [{\citenamefont {Wang}\ \emph {et~al.}(2019)\citenamefont {Wang},
  \citenamefont {Rigg}, \citenamefont {Sethian}, \citenamefont {Sinclair},
  \citenamefont {Weir}, \citenamefont {Williams}, \citenamefont {Zhang},
  \citenamefont {Hawreliak}, \citenamefont {Toyoda}, \citenamefont {Gupta},
  \citenamefont {Li}, \citenamefont {Broege}, \citenamefont {Bromage},
  \citenamefont {Earley}, \citenamefont {Guy},\ and\ \citenamefont
  {Zuegel}}]{Wang2019}%
  \BibitemOpen
  \bibfield  {author} {\bibinfo {author} {\bibfnamefont {X.}~\bibnamefont
  {Wang}}, \bibinfo {author} {\bibfnamefont {P.}~\bibnamefont {Rigg}}, \bibinfo
  {author} {\bibfnamefont {J.}~\bibnamefont {Sethian}}, \bibinfo {author}
  {\bibfnamefont {N.}~\bibnamefont {Sinclair}}, \bibinfo {author}
  {\bibfnamefont {N.}~\bibnamefont {Weir}}, \bibinfo {author} {\bibfnamefont
  {B.}~\bibnamefont {Williams}}, \bibinfo {author} {\bibfnamefont
  {J.}~\bibnamefont {Zhang}}, \bibinfo {author} {\bibfnamefont
  {J.}~\bibnamefont {Hawreliak}}, \bibinfo {author} {\bibfnamefont
  {Y.}~\bibnamefont {Toyoda}}, \bibinfo {author} {\bibfnamefont
  {Y.}~\bibnamefont {Gupta}}, \bibinfo {author} {\bibfnamefont
  {Y.}~\bibnamefont {Li}}, \bibinfo {author} {\bibfnamefont {D.}~\bibnamefont
  {Broege}}, \bibinfo {author} {\bibfnamefont {J.}~\bibnamefont {Bromage}},
  \bibinfo {author} {\bibfnamefont {R.}~\bibnamefont {Earley}}, \bibinfo
  {author} {\bibfnamefont {D.}~\bibnamefont {Guy}}, \ and\ \bibinfo {author}
  {\bibfnamefont {J.}~\bibnamefont {Zuegel}},\ }\href {\doibase
  10.1063/1.5088367} {\bibfield  {journal} {\bibinfo  {journal} {Review of
  Scientific Instruments}\ }\textbf {\bibinfo {volume} {90}},\ \bibinfo {pages}
  {053901} (\bibinfo {year} {2019})},\ \Eprint
  {http://arxiv.org/abs/https://doi.org/10.1063/1.5088367}
  {https://doi.org/10.1063/1.5088367} \BibitemShut {NoStop}%
\bibitem [{\citenamefont {Jensen}\ \emph {et~al.}(2014)\citenamefont {Jensen},
  \citenamefont {Ramos}, \citenamefont {Iverson}, \citenamefont {Bernier},
  \citenamefont {Carlson}, \citenamefont {Yeager}, \citenamefont {Fezzaa},\
  and\ \citenamefont {Hooks}}]{Jensen2014}%
  \BibitemOpen
  \bibfield  {author} {\bibinfo {author} {\bibfnamefont {B.~J.}\ \bibnamefont
  {Jensen}}, \bibinfo {author} {\bibfnamefont {K.~J.}\ \bibnamefont {Ramos}},
  \bibinfo {author} {\bibfnamefont {A.~J.}\ \bibnamefont {Iverson}}, \bibinfo
  {author} {\bibfnamefont {J.}~\bibnamefont {Bernier}}, \bibinfo {author}
  {\bibfnamefont {C.~A.}\ \bibnamefont {Carlson}}, \bibinfo {author}
  {\bibfnamefont {J.~D.}\ \bibnamefont {Yeager}}, \bibinfo {author}
  {\bibfnamefont {K.}~\bibnamefont {Fezzaa}}, \ and\ \bibinfo {author}
  {\bibfnamefont {D.~E.}\ \bibnamefont {Hooks}},\ }\href {\doibase
  10.1088/1742-6596/500/4/042001} {\bibfield  {journal} {\bibinfo  {journal}
  {J. Phys. Conf. Ser.}\ }\textbf {\bibinfo {volume} {500}},\ \bibinfo {pages}
  {042001} (\bibinfo {year} {2014})}\BibitemShut {NoStop}%
\bibitem [{\citenamefont {{Gupta}}\ \emph {et~al.}(2012)\citenamefont
  {{Gupta}}, \citenamefont {{Turneaure}}, \citenamefont {{Perkins}},
  \citenamefont {{Zimmerman}}, \citenamefont {{Arganbright}}, \citenamefont
  {{Shen}},\ and\ \citenamefont {{Chow}}}]{Gupta2012}%
  \BibitemOpen
  \bibfield  {author} {\bibinfo {author} {\bibfnamefont {Y.~M.}\ \bibnamefont
  {{Gupta}}}, \bibinfo {author} {\bibfnamefont {S.~J.}\ \bibnamefont
  {{Turneaure}}}, \bibinfo {author} {\bibfnamefont {K.}~\bibnamefont
  {{Perkins}}}, \bibinfo {author} {\bibfnamefont {K.}~\bibnamefont
  {{Zimmerman}}}, \bibinfo {author} {\bibfnamefont {N.}~\bibnamefont
  {{Arganbright}}}, \bibinfo {author} {\bibfnamefont {G.}~\bibnamefont
  {{Shen}}}, \ and\ \bibinfo {author} {\bibfnamefont {P.}~\bibnamefont
  {{Chow}}},\ }\href {\doibase 10.1063/1.4772577} {\bibfield  {journal}
  {\bibinfo  {journal} {Rev. Sci. Instrum.}\ }\textbf {\bibinfo {volume}
  {83}},\ \bibinfo {eid} {123905-123905-10} (\bibinfo {year}
  {2012})}\BibitemShut {NoStop}%
\bibitem [{\citenamefont {Turneaure}\ \emph {et~al.}(2016)\citenamefont
  {Turneaure}, \citenamefont {Sinclair},\ and\ \citenamefont
  {Gupta}}]{Turneaure2016}%
  \BibitemOpen
  \bibfield  {author} {\bibinfo {author} {\bibfnamefont {S.~J.}\ \bibnamefont
  {Turneaure}}, \bibinfo {author} {\bibfnamefont {N.}~\bibnamefont {Sinclair}},
  \ and\ \bibinfo {author} {\bibfnamefont {Y.~M.}\ \bibnamefont {Gupta}},\
  }\href {\doibase 10.1103/PhysRevLett.117.045502} {\bibfield  {journal}
  {\bibinfo  {journal} {Phys. Rev. Lett.}\ }\textbf {\bibinfo {volume} {117}},\
  \bibinfo {pages} {045502} (\bibinfo {year} {2016})}\BibitemShut {NoStop}%
\bibitem [{\citenamefont {Prescher}\ and\ \citenamefont
  {Prakapenka}(2015)}]{Prescher2015}%
  \BibitemOpen
  \bibfield  {author} {\bibinfo {author} {\bibfnamefont {C.}~\bibnamefont
  {Prescher}}\ and\ \bibinfo {author} {\bibfnamefont {V.~B.}\ \bibnamefont
  {Prakapenka}},\ }\href@noop {} {\bibfield  {journal} {\bibinfo  {journal}
  {High Pres. Res.}\ }\textbf {\bibinfo {volume} {35}},\ \bibinfo {pages} {223}
  (\bibinfo {year} {2015})}\BibitemShut {NoStop}%
\bibitem [{\citenamefont {Stavrou}\ \emph {et~al.}(2020)\citenamefont
  {Stavrou}, \citenamefont {Bagge-Hansen}, \citenamefont {Hammons},
  \citenamefont {Nielsen}, \citenamefont {Steele}, \citenamefont {Xiao},
  \citenamefont {Kroonblawd}, \citenamefont {Nelms}, \citenamefont {Shaw},
  \citenamefont {Bassett}, \citenamefont {Bastea}, \citenamefont {Lauderbach},
  \citenamefont {Hodgin}, \citenamefont {Perez-Marty}, \citenamefont {Singh},
  \citenamefont {Das}, \citenamefont {Li}, \citenamefont {Schuman},
  \citenamefont {Sinclair}, \citenamefont {Fezzaa}, \citenamefont {Deriy},
  \citenamefont {Leininger},\ and\ \citenamefont {Willey}}]{Stavrou2020}%
  \BibitemOpen
  \bibfield  {author} {\bibinfo {author} {\bibfnamefont {E.}~\bibnamefont
  {Stavrou}}, \bibinfo {author} {\bibfnamefont {M.}~\bibnamefont
  {Bagge-Hansen}}, \bibinfo {author} {\bibfnamefont {J.~A.}\ \bibnamefont
  {Hammons}}, \bibinfo {author} {\bibfnamefont {M.~H.}\ \bibnamefont
  {Nielsen}}, \bibinfo {author} {\bibfnamefont {B.~A.}\ \bibnamefont {Steele}},
  \bibinfo {author} {\bibfnamefont {P.}~\bibnamefont {Xiao}}, \bibinfo {author}
  {\bibfnamefont {M.~P.}\ \bibnamefont {Kroonblawd}}, \bibinfo {author}
  {\bibfnamefont {M.~D.}\ \bibnamefont {Nelms}}, \bibinfo {author}
  {\bibfnamefont {W.~L.}\ \bibnamefont {Shaw}}, \bibinfo {author}
  {\bibfnamefont {W.}~\bibnamefont {Bassett}}, \bibinfo {author} {\bibfnamefont
  {S.}~\bibnamefont {Bastea}}, \bibinfo {author} {\bibfnamefont {L.~M.}\
  \bibnamefont {Lauderbach}}, \bibinfo {author} {\bibfnamefont {R.~L.}\
  \bibnamefont {Hodgin}}, \bibinfo {author} {\bibfnamefont {N.~A.}\
  \bibnamefont {Perez-Marty}}, \bibinfo {author} {\bibfnamefont
  {S.}~\bibnamefont {Singh}}, \bibinfo {author} {\bibfnamefont
  {P.}~\bibnamefont {Das}}, \bibinfo {author} {\bibfnamefont {Y.}~\bibnamefont
  {Li}}, \bibinfo {author} {\bibfnamefont {A.}~\bibnamefont {Schuman}},
  \bibinfo {author} {\bibfnamefont {N.}~\bibnamefont {Sinclair}}, \bibinfo
  {author} {\bibfnamefont {K.}~\bibnamefont {Fezzaa}}, \bibinfo {author}
  {\bibfnamefont {A.}~\bibnamefont {Deriy}}, \bibinfo {author} {\bibfnamefont
  {L.~D.}\ \bibnamefont {Leininger}}, \ and\ \bibinfo {author} {\bibfnamefont
  {T.~M.}\ \bibnamefont {Willey}},\ }\href {\doibase
  10.1103/PhysRevB.102.104116} {\bibfield  {journal} {\bibinfo  {journal}
  {Phys. Rev. B}\ }\textbf {\bibinfo {volume} {102}},\ \bibinfo {pages}
  {104116} (\bibinfo {year} {2020})}\BibitemShut {NoStop}%
\bibitem [{\citenamefont {Kraus}\ and\ \citenamefont
  {Nolze}(1996)}]{Kraus1996}%
  \BibitemOpen
  \bibfield  {author} {\bibinfo {author} {\bibfnamefont {W.}~\bibnamefont
  {Kraus}}\ and\ \bibinfo {author} {\bibfnamefont {G.}~\bibnamefont {Nolze}},\
  }\href@noop {} {\bibfield  {journal} {\bibinfo  {journal} {J. Appl.
  Crystallogr.}\ }\textbf {\bibinfo {volume} {29}},\ \bibinfo {pages} {301}
  (\bibinfo {year} {1996})}\BibitemShut {NoStop}%
\bibitem [{\citenamefont {Boultif}\ and\ \citenamefont
  {Lou\"{e}r}(2004)}]{Boutlif2004}%
  \BibitemOpen
  \bibfield  {author} {\bibinfo {author} {\bibfnamefont {A.}~\bibnamefont
  {Boultif}}\ and\ \bibinfo {author} {\bibfnamefont {D.}~\bibnamefont
  {Lou\"{e}r}},\ }\href@noop {} {\bibfield  {journal} {\bibinfo  {journal} {J.
  Appl. Crystallogr.}\ }\textbf {\bibinfo {volume} {37}},\ \bibinfo {pages}
  {724} (\bibinfo {year} {2004})}\BibitemShut {NoStop}%
\bibitem [{sup()}]{supp}%
  \BibitemOpen
  \href@noop {} {}\bibinfo {howpublished} {See Supplemental Material at , which
  includes supplemental figures 1-3}\BibitemShut {NoStop}%
\bibitem [{\citenamefont {Cady}\ and\ \citenamefont {Larson}(1965)}]{Cady1965}%
  \BibitemOpen
  \bibfield  {author} {\bibinfo {author} {\bibfnamefont {H.}~\bibnamefont
  {Cady}}\ and\ \bibinfo {author} {\bibfnamefont {A.}~\bibnamefont {Larson}},\
  }\href {\doibase {10.1107/S0365110X6500107X}} {\bibfield  {journal} {\bibinfo
   {journal} {Acta Crystallogr.}\ }\textbf {\bibinfo {volume} {18}},\ \bibinfo
  {pages} {485} (\bibinfo {year} {1965})}\BibitemShut {NoStop}%
\bibitem [{\citenamefont {Hamilton}\ \emph {et~al.}(2019)\citenamefont
  {Hamilton}, \citenamefont {Kroonblawd}, \citenamefont {Islam},\ and\
  \citenamefont {Strachan}}]{Hamilton2019}%
  \BibitemOpen
  \bibfield  {author} {\bibinfo {author} {\bibfnamefont {B.}~\bibnamefont
  {Hamilton}}, \bibinfo {author} {\bibfnamefont {M.~P.}\ \bibnamefont
  {Kroonblawd}}, \bibinfo {author} {\bibfnamefont {M.~M.}\ \bibnamefont
  {Islam}}, \ and\ \bibinfo {author} {\bibfnamefont {A.}~\bibnamefont
  {Strachan}},\ }\href {\doibase 10.1021/acs.jpcc.9b05409} {\bibfield
  {journal} {\bibinfo  {journal} {J. Phys. Chem. C}\ }\textbf {\bibinfo
  {volume} {123}},\ \bibinfo {pages} {21969} (\bibinfo {year}
  {2019})}\BibitemShut {NoStop}%
\bibitem [{\citenamefont {Leversee}\ \emph {et~al.}(2019)\citenamefont
  {Leversee}, \citenamefont {Zaug}, \citenamefont {Sain}, \citenamefont {Weir},
  \citenamefont {Bastea}, \citenamefont {Fried},\ and\ \citenamefont
  {Stavrou}}]{Leversee2019}%
  \BibitemOpen
  \bibfield  {author} {\bibinfo {author} {\bibfnamefont {R.~A.}\ \bibnamefont
  {Leversee}}, \bibinfo {author} {\bibfnamefont {J.~M.}\ \bibnamefont {Zaug}},
  \bibinfo {author} {\bibfnamefont {J.~D.}\ \bibnamefont {Sain}}, \bibinfo
  {author} {\bibfnamefont {S.~T.}\ \bibnamefont {Weir}}, \bibinfo {author}
  {\bibfnamefont {S.}~\bibnamefont {Bastea}}, \bibinfo {author} {\bibfnamefont
  {L.~E.}\ \bibnamefont {Fried}}, \ and\ \bibinfo {author} {\bibfnamefont
  {E.}~\bibnamefont {Stavrou}},\ }\href {\doibase 10.1063/1.5108677} {\bibfield
   {journal} {\bibinfo  {journal} {Applied Physics Letters}\ }\textbf {\bibinfo
  {volume} {115}},\ \bibinfo {pages} {051902} (\bibinfo {year}
  {2019})}\BibitemShut {NoStop}%
\bibitem [{\citenamefont {Degtyarev}\ \emph {et~al.}(2016)\citenamefont
  {Degtyarev}, \citenamefont {Smirnov}, \citenamefont {Kostitsin},
  \citenamefont {Stankevich}, \citenamefont {Muzyrya}, \citenamefont {Ten},
  \citenamefont {Pruuel}, \citenamefont {Kashkarov},\ and\ \citenamefont
  {Batretdinova}}]{Degtyarev2016}%
  \BibitemOpen
  \bibfield  {author} {\bibinfo {author} {\bibfnamefont {A.~A.}\ \bibnamefont
  {Degtyarev}}, \bibinfo {author} {\bibfnamefont {E.~B.}\ \bibnamefont
  {Smirnov}}, \bibinfo {author} {\bibfnamefont {O.~V.}\ \bibnamefont
  {Kostitsin}}, \bibinfo {author} {\bibfnamefont {A.~V.}\ \bibnamefont
  {Stankevich}}, \bibinfo {author} {\bibfnamefont {A.~K.}\ \bibnamefont
  {Muzyrya}}, \bibinfo {author} {\bibfnamefont {K.~A.}\ \bibnamefont {Ten}},
  \bibinfo {author} {\bibfnamefont {E.~R.}\ \bibnamefont {Pruuel}}, \bibinfo
  {author} {\bibfnamefont {A.~O.}\ \bibnamefont {Kashkarov}}, \ and\ \bibinfo
  {author} {\bibfnamefont {L.~H.}\ \bibnamefont {Batretdinova}},\ }\href
  {http://stacks.iop.org/1742-6596/774/i=1/a=012011} {\bibfield  {journal}
  {\bibinfo  {journal} {J. Phys. Conf. Ser.}\ }\textbf {\bibinfo {volume}
  {774}},\ \bibinfo {pages} {012011} (\bibinfo {year} {2016})}\BibitemShut
  {NoStop}%
\bibitem [{\citenamefont {Gilardi}\ and\ \citenamefont
  {Butcher}(2001)}]{Gilardi2001}%
  \BibitemOpen
  \bibfield  {author} {\bibinfo {author} {\bibfnamefont {R.~D.}\ \bibnamefont
  {Gilardi}}\ and\ \bibinfo {author} {\bibfnamefont {R.~J.}\ \bibnamefont
  {Butcher}},\ }\href {\doibase 10.1107/S1600536801010352} {\bibfield
  {journal} {\bibinfo  {journal} {Acta Crystallogr. E}\ }\textbf {\bibinfo
  {volume} {57}},\ \bibinfo {pages} {o657} (\bibinfo {year}
  {2001})}\BibitemShut {NoStop}%
\bibitem [{\citenamefont {Field}(1992)}]{Field1992}%
  \BibitemOpen
  \bibfield  {author} {\bibinfo {author} {\bibfnamefont {J.~E.}\ \bibnamefont
  {Field}},\ }\href {\doibase 10.1021/ar00023a002} {\bibfield  {journal}
  {\bibinfo  {journal} {Acc. Chem. Res.}\ }\textbf {\bibinfo {volume} {25}},\
  \bibinfo {pages} {489} (\bibinfo {year} {1992})}\BibitemShut {NoStop}%
\bibitem [{\citenamefont {Campbell}\ \emph {et~al.}(1961)\citenamefont
  {Campbell}, \citenamefont {Davis}, \citenamefont {Ramsay},\ and\
  \citenamefont {Travis}}]{Campbell1961}%
  \BibitemOpen
  \bibfield  {author} {\bibinfo {author} {\bibfnamefont {A.~W.}\ \bibnamefont
  {Campbell}}, \bibinfo {author} {\bibfnamefont {W.~C.}\ \bibnamefont {Davis}},
  \bibinfo {author} {\bibfnamefont {J.~B.}\ \bibnamefont {Ramsay}}, \ and\
  \bibinfo {author} {\bibfnamefont {J.~R.}\ \bibnamefont {Travis}},\ }\href
  {\doibase 10.1063/1.1706354} {\bibfield  {journal} {\bibinfo  {journal}
  {Phys. Fluids}\ }\textbf {\bibinfo {volume} {4}},\ \bibinfo {pages} {511}
  (\bibinfo {year} {1961})}\BibitemShut {NoStop}%
\bibitem [{\citenamefont {F. P.~Bowden}(1952)}]{Bowden1952}%
  \BibitemOpen
  \bibfield  {author} {\bibinfo {author} {\bibfnamefont {A.~Y.}\ \bibnamefont
  {F. P.~Bowden}},\ }\href {\doibase 10.1017/S0368393100125064} {\bibfield
  {journal} {\bibinfo  {journal} {RAeS}\ }\textbf {\bibinfo {volume} {56}},\
  \bibinfo {pages} {807–808} (\bibinfo {year} {1952})}\BibitemShut {NoStop}%
\bibitem [{\citenamefont {Kirkwood}\ and\ \citenamefont
  {Wood}(1954)}]{Kirkwood1954}%
  \BibitemOpen
  \bibfield  {author} {\bibinfo {author} {\bibfnamefont {J.~G.}\ \bibnamefont
  {Kirkwood}}\ and\ \bibinfo {author} {\bibfnamefont {W.~W.}\ \bibnamefont
  {Wood}},\ }\href {\doibase 10.1063/1.1739939} {\bibfield  {journal} {\bibinfo
   {journal} {The Journal of Chemical Physics}\ }\textbf {\bibinfo {volume}
  {22}},\ \bibinfo {pages} {1915} (\bibinfo {year} {1954})}\BibitemShut
  {NoStop}%
\bibitem [{\citenamefont {Loboiko}\ and\ \citenamefont
  {Lubyatinsky}(2000)}]{Loboiko2000}%
  \BibitemOpen
  \bibfield  {author} {\bibinfo {author} {\bibfnamefont {B.~G.}\ \bibnamefont
  {Loboiko}}\ and\ \bibinfo {author} {\bibfnamefont {S.~N.}\ \bibnamefont
  {Lubyatinsky}},\ }\href {https://doi.org/10.1023/A:1002898505288} {\bibfield
  {journal} {\bibinfo  {journal} {Combustion, Explosion and Shock Waves}\
  }\textbf {\bibinfo {volume} {36}},\ \bibinfo {pages} {716} (\bibinfo {year}
  {2000})}\BibitemShut {NoStop}%
\bibitem [{\citenamefont {Manaa}\ \emph {et~al.}(2009)\citenamefont {Manaa},
  \citenamefont {Reed}, \citenamefont {Fried},\ and\ \citenamefont
  {Goldman}}]{Manaa2009}%
  \BibitemOpen
  \bibfield  {author} {\bibinfo {author} {\bibfnamefont {M.~R.}\ \bibnamefont
  {Manaa}}, \bibinfo {author} {\bibfnamefont {E.~J.}\ \bibnamefont {Reed}},
  \bibinfo {author} {\bibfnamefont {L.~E.}\ \bibnamefont {Fried}}, \ and\
  \bibinfo {author} {\bibfnamefont {N.}~\bibnamefont {Goldman}},\ }\href
  {\doibase 10.1021/ja808196e} {\bibfield  {journal} {\bibinfo  {journal} {J.
  Am. Chem. Soc.}\ }\textbf {\bibinfo {volume} {131}},\ \bibinfo {pages} {5483}
  (\bibinfo {year} {2009})}\BibitemShut {NoStop}%
\end{thebibliography}
\end{document}